\documentclass{article}

\usepackage[preprint]{neurips_2026}

\makeatletter
\renewcommand*{\@fnsymbol}[1]{%
  \ifcase#1
  \or \ensuremath{\dagger}
  \or \ensuremath{*}
  \else \@ctrerr
  \fi
}
\makeatother

\setcitestyle{authoryear,open={(},close={)},semicolon,aysep={,\,}}

\usepackage[utf8]{inputenc} 
\usepackage[T1]{fontenc}    
\usepackage{hyperref}       
\usepackage{url}            
\usepackage{booktabs}       
\usepackage{amsfonts}       
\usepackage{nicefrac}       
\usepackage{microtype}      
\usepackage{xcolor}         

\usepackage{graphicx}
\usepackage{framed,multirow}
\usepackage{amssymb}
\usepackage{amsmath}
\usepackage{latexsym}
\usepackage{color}
\usepackage{colortbl}
\usepackage{epstopdf}
\usepackage{caption}
\usepackage{epsfig}
\usepackage{array}
\usepackage{epsfig}
\usepackage{comment}
\usepackage{tabularx}
\usepackage{makecell}
\usepackage{multirow}
\usepackage{url}
\usepackage{pifont}
\usepackage{enumitem}
\usepackage{romannum}

\usepackage{listings}
\usepackage[ruled,vlined]{algorithm2e}

\usepackage{fancyhdr}       

\hypersetup{hidelinks}

\newcommand{\RNum}[1]{\uppercase\expandafter{\romannumeral #1\relax}}
\newcommand{\Mten}{\mathbf{M}}
\newcommand{\Sten}{\mathbf{\Sigma}}
\newcommand{\Xten}{\mathbf{X}}
\newcommand{\Yten}{\mathbf{Y}}
\newcommand{\Zten}{\mathbf{Z}}
\newcommand{\Lten}{\mathbf{L}}
\newcommand{\Uten}{\mathbf{U}}
\newcommand{\Cten}{\mathbf{C}}
\newcommand{\Dten}{\mathbf{D}}

\newcommand{\ie}{\textit{i.e.},\xspace}

\title{Bound-Conditioned Latent Inference for Progressive Image Compression}

\author{
  Jaeseok Jang\thanks{Jaeseok Jang and Seungmin Jeon contributed equally to this work and are co-first authors.}\\
  Korea University \\
  \texttt{jsjang@mcl.korea.ac.kr} \\
  \And
  Seungmin Jeon\footnotemark[1] \\
  Korea University \\
  \texttt{seungminjeon@mcl.korea.ac.kr} \\
  \And
  Kwang Pyo Choi \\
  Samsung Electronics \\
  \texttt{kp5.choi@samsung.com} \\
  \And
  Chang-Su Kim\thanks{Corresponding author: Chang-Su Kim, <changsukim@korea.ac.kr>} \\
  Korea University \\
  \texttt{changsukim@korea.ac.kr}
}

\begin{document}

\maketitle
\pagenumbering{arabic}

\newcolumntype{L}[1]{>{\raggedright\arraybackslash}p{#1}}
\newcolumntype{C}[1]{>{\centering\arraybackslash}p{#1}}
\newcolumntype{R}[1]{>{\raggedleft\arraybackslash}p{#1}}

\begin{abstract}
Progressive image compression reconstructs images from prefixes of a single bitstream and therefore involves sequential latent inference under partial observations. Existing trit-plane codecs mainly treat decoded trits as discrete symbols for probability prediction, without fully exploiting their role as interval constraints on the underlying continuous latents. We propose bound-conditioned latent inference (BLI), a framework that reformulates progressive trit-plane coding as sequential latent inference under shrinking interval constraints. At each coding step, the trits decoded so far define lower and upper bounds for each latent element. BLI uses the center and width of these bounds, together with hyperprior information shared by the encoder and decoder, to jointly refine the Gaussian parameters $(\mu,\sigma)$ for entropy modeling and the latent estimate $\hat{y}$ for reconstruction. As additional trits are decoded, the feasible intervals shrink monotonically, yielding tighter constraints for the next inference step and enabling flexible in-plane refinement schedules. Experiments on Kodak, CLIC, and JPEG-AI show that BLI achieves state-of-the-art rate-distortion performance among progressive image codecs, reducing BD-rate by $17.78\%$ over DPICT and by $7.96\%$ over CTC on Kodak, while replacing CTC's plane-specific predictors with a shared model that uses about $5.3\times$ fewer parameters and achieves lower latency than CTC.
\end{abstract}

\section{Introduction}
\label{sec:intro}

Progressive image compression differs from standard fixed-rate compression in that it allows the decoder to reconstruct an image from prefixes of a single bitstream. At an intermediate decoding step, only part of the latent representation has been transmitted, while the unresolved latent values must be inferred from partial observations. Thus, progressive compression is not merely a rate-distortion (R-D) optimization problem, but also a sequential latent inference problem under partial observations. This property is essential for adaptive streaming, bandwidth-limited transmission, and storage systems that require multiple quality levels from a single bitstream.

Traditional and learned codecs have achieved strong R-D performance, from hand-crafted transform codecs such as JPEG and JPEG2000 \citep{y1992_CE_JPEG,y2001_SPM_JPEG2000} to neural codecs with learned transforms, hyperpriors, and context models \citep{y2017_NIPS_agustsson,y2017_ICLR_balle,y2018_ICLR_balle,y2018_NIPS_minnen,y2021_CVPR_he}. However, most high-performance learned codecs are optimized for a single target rate. Multirate codecs \citep{y2021_CVPR_cui,y2021_CVPR_yang,y2025_CVPR_xu} improve rate flexibility, but generally require separately encoded bitstreams for different rates, in contrast to progressive codecs, every bitstream prefix of which remains decodable.

Learned progressive codecs provide decodable bitstream prefixes by decomposing latent representations into successively refinable components. PLONQ \citep{y2021_ICIP_lu} uses nested quantization grids for coarse-to-fine refinement, while DPICT \citep{y2022_CVPR_lee} discretizes each latent element into trits that are transmitted sequentially. CTC \citep{y2023_CVPR_jeon} further improves trit-plane coding by introducing context models that refine ternary probabilities and partial reconstructions. However, existing trit-plane codecs still primarily treat decoded trits as discrete symbols for probability prediction and reconstruction update. In particular, CTC improves discrete trit prediction and partial reconstruction, but it does not explicitly perform in-plane continuous latent inference conditioned on the feasible intervals induced by previously decoded trits. The key observation behind our approach is that decoded trits are not merely discrete symbols: in trit-plane coding, they also restrict each underlying continuous latent element to a feasible interval, which can serve as an explicit constraint for latent inference.

\begin{figure*}[t]
\vspace*{-0.3cm}
\centering
\includegraphics[width=\linewidth]{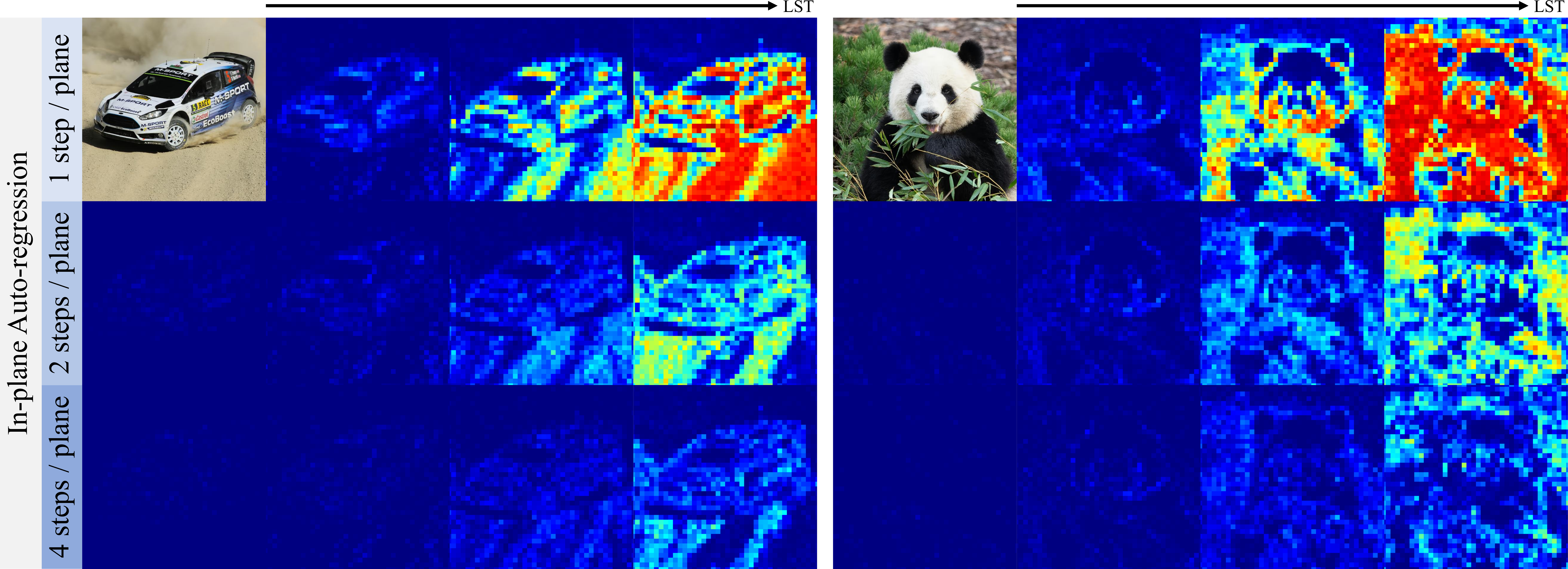}
\caption{
Rate-allocation maps of the proposed bound-conditioned latent inference (BLI) codec. Each column corresponds to the same trit-plane level, \ie the same degree of latent quantization, while each row uses a different number of in-plane refinement steps. The top row encodes an entire trit-plane in a single step, whereas the middle and bottom rows divide each trit-plane into two and four substeps, respectively. Colors indicate spatial rate allocation, with warmer colors corresponding to higher coding rates and cooler colors to lower rates. More frequent bound-conditioned updates iteratively refine the latent distribution within the same quantization level, yielding better rate efficiency without changing the final trit-plane precision.
}
\vspace*{-0.4cm}
\label{fig:intro}
\end{figure*}

We present bound-conditioned latent inference (BLI), a framework that explicitly exploits these feasible intervals by reformulating progressive trit-plane coding as sequential latent inference under shrinking interval constraints. In BLI, each partially decoded latent element is modeled as a continuous variable constrained by lower and upper bounds determined by the trits decoded so far. At each coding step, BLI takes the center and width of the current feasible interval, together with hyperprior information shared by the encoder and decoder, and jointly refines the Gaussian parameters $(\mu,\sigma)$ for entropy modeling and the latent estimate $\hat{y}$ for reconstruction. As additional trits are decoded, the feasible intervals shrink monotonically, yielding tighter constraints for the next inference step. As illustrated in Figure~\ref{fig:intro}, this bound-conditioned formulation naturally supports flexible in-plane transmission schedules: each trit-plane can be encoded at once or divided into finer substeps, allowing the model to update the latent distribution and latent estimate more frequently. Conceptually, BLI can be interpreted as constraint-consistent latent refinement, related to projection-based feasibility methods such as POCS \citep{y1996_SIAM_Bauschke} and to denoising or inpainting under interval constraints. Unlike exact metric projection, however, BLI uses a learned amortized estimator trained directly for entropy modeling and reconstruction quality.

Our main contributions are summarized as follows:

\begin{itemize}[leftmargin=2.0em,itemsep=0.1em,topsep=0.1em]
\item We reformulate progressive trit-plane coding as sequential latent inference under shrinking interval constraints, where decoded trits define feasible ranges for the underlying continuous latent elements.

\item We propose BLI, a learned constrained inference framework that jointly refines entropy-model parameters and latent estimates from bound and hyperprior information, with flexible in-plane autoregressive updates within each trit-plane.

\item Experiments show that BLI achieves state-of-the-art R\mbox{-}D performance among progressive image codecs, reducing BD-rate by $17.78\%$ over DPICT and by $7.96\%$ over CTC on the Kodak dataset, while replacing CTC's plane-specific prediction networks with a shared model that uses about $5.3\times$ fewer parameters and achieves lower latency than CTC.
\end{itemize}

\section{Related Work}
\label{sec:related}

\noindent\textbf{Learned image compression:}
Traditional codecs \citep{y1992_CE_JPEG,bpg,y2021_TCSVT_Bross} follow the transform coding paradigm \citep{y2001_spm_goyal}, while learned codecs replace hand-crafted transforms and entropy models with neural networks. Early learned codecs introduced differentiable quantization and end-to-end R-D optimization \citep{y2017_NIPS_agustsson,y2017_ICLR_balle}, and hyperprior-based models improved entropy estimation by transmitting side information about latent distributions \citep{y2018_ICLR_balle}. Later methods further developed expressive priors, including Gaussian mixture and asymmetric Gaussian models \citep{y2020_CVPR_cheng,y2021_CVPR_cui}.

Entropy models improve probability estimation by exploiting dependencies among latent elements. Autoregressive context models \citep{y2018_NIPS_minnen,y2018_CVPR_mentzer,y2019_ICLR_lee} improve compression efficiency through sequential prediction, but incur high coding latency due to limited parallelism. Partially parallel models \citep{y2020_ICIP_minnen,y2021_CVPR_he} reduce this cost by exploiting latent correlations more efficiently.

Recent learned codecs incorporate attention mechanisms and transformers into analysis-synthesis networks and entropy models. Transformer-based codecs \citep{y2022_ICLR_zhu,y2022_CVPR_zou,y2023_CVPR_liu,y2024_ICLR_li} improve global context modeling and yield compact latents, while transformer-based entropy models \citep{y2022_ICLR_qian,y2022_CVPR_kim} capture long-range dependencies for more accurate probability estimation. These advances provide strong fixed-rate baselines, but most of them are not designed to produce a single progressively decodable bitstream.

\medskip
\noindent\textbf{Scalable image compression:}
Beyond fixed-rate learned codecs, scalable compression aims to support multiple bitrates using a single trained model or a single embedded bitstream. Variable-rate methods adjust the operating rate by modulating internal representations, such as gain parameters, channel width, or spatial transforms \citep{y2021_CVPR_cui,y2021_CVPR_yang,y2021_ICCV_song}, or by using adjustable quantizers \citep{y2020_ICASSP_chen,y2023_ICIP_tong,y2025_CVPR_xu}. Although these methods provide flexible rate control from a single model, each target bitrate generally requires a separately encoded bitstream. Therefore, they differ from progressive codecs, in which every prefix of a single bitstream remains decodable at multiple quality levels.

Progressive image compression uses embedded bitstreams, where reconstruction quality improves as additional bits are appended. Classical codecs such as progressive JPEG \citep{y1992_CE_JPEG}, JPEG2000 \citep{y2001_SPM_JPEG2000}, and SPIHT \citep{y1996_TCSVT_said} achieve this through spectral selection, embedded bitstreams, or significance-driven coefficient ordering. In learned compression, PLONQ \citep{y2021_ICIP_lu} introduced nested quantization for coarse-to-fine refinement, while DPICT \citep{y2022_CVPR_lee} realized fine-grained scalability by slicing latents into trit-planes. CTC \citep{y2023_CVPR_jeon} further improved ternary probability modeling and partial reconstructions. However, DPICT and CTC still primarily model decoded trits as discrete symbols. As a result, the feasible intervals induced by decoded trits are not explicitly exploited for in-plane continuous latent inference, which is the key limitation addressed by BLI.

\section{Proposed Algorithm}
\label{sec:proposed}

We present bound-conditioned latent inference (BLI) for progressive trit-plane coding. BLI treats each partially decoded latent element as a continuous variable constrained by lower and upper bounds determined by the trits decoded so far. Within these bounds, BLI jointly refines the Gaussian parameters $(\mu,\sigma)$ for entropy modeling and the latent estimate $\hat{y}$ for reconstruction as the bounds shrink over coding steps. Unlike CTC \citep{y2023_CVPR_jeon}, which improves ternary probability modeling and partial reconstruction through discrete context prediction, BLI performs bound-conditioned Gaussian inference at every refinement step. This enables iterative in-plane refinement of both the latent distribution and latent estimate, improving entropy modeling accuracy and reconstruction quality.

\subsection{Bound-Conditioned Latent Inference}
\label{ssec:formulation}

\begin{figure*}[t]
\centering
\includegraphics[width=\linewidth]{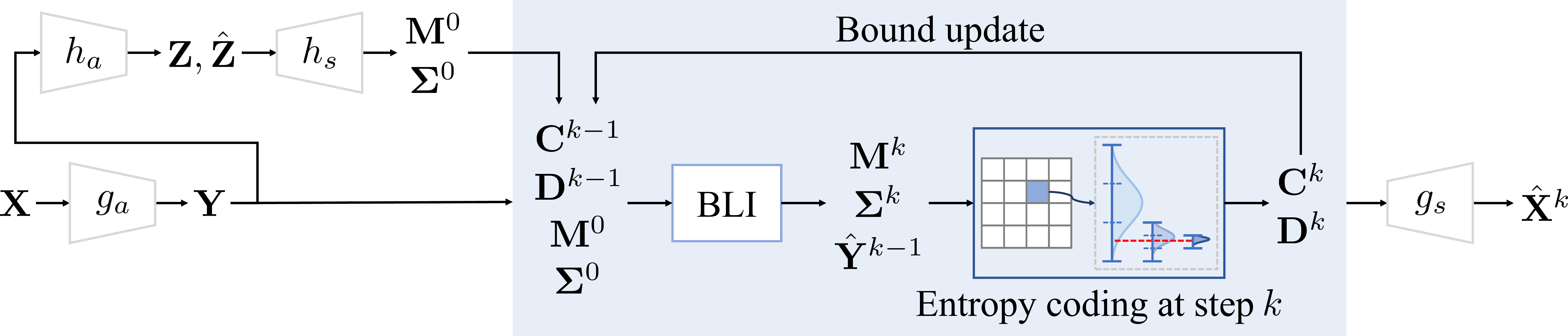}
\setlength{\abovecaptionskip}{-0.1cm}
\caption{
Overview of the BLI process. At coding step $k$, the trits decoded so far define bound tensors $(\Cten^{k-1}, \Dten^{k-1})$, which represent the current feasible ranges of latent elements. Given these bounds and the hyperprior parameters $(\Mten^0,\Sten^0)$, BLI estimates the Gaussian parameters $(\Mten^k,\Sten^k)$ for entropy coding and the latent tensor $\hat{\Yten}^{k-1}$ for reconstruction under the current bounds. After a subset of trits is decoded using $(\Mten^k,\Sten^k)$, the bounds are updated to $(\Cten^k,\Dten^k)$, yielding tighter feasible ranges that, in turn, are used in the next coding step. This step-wise process iteratively refines the latent distribution and reconstruction.
}
\label{fig:overview}
\end{figure*}

Suppose that an image $\Xten$ is transformed into a latent tensor $\Yten\in\mathbb{R}^{C\times H\times W}$, as illustrated in Figure~\ref{fig:overview}. Each element $y_i$ in $\Yten$ is represented by trits and progressively transmitted. BLI proceeds in coding steps indexed by a global counter $k=1,2,\cdots$, which increases monotonically across trit-planes. At step $k$, we define the following state variables.
\begin{itemize}[leftmargin=2.0em,itemsep=0.1em,topsep=0.1em]
  \item $(\Mten^k, \Sten^k)$: Gaussian mean and standard deviation parameters used for entropy coding at step $k$.
  \item $\hat{\Yten}^{k}$: Reconstructed latent estimate consistent with the bounds after step $k$.
  \item $(\Lten^{k}, \Uten^{k})$: Lower and upper bounds of the latent elements after step $k$.
  \item $(\Cten^{k}, \Dten^{k})$: Centers and widths of the feasible intervals after step $k$, defined as
  \begin{equation}
  \Cten^{k}=\frac{\Lten^{k}+\Uten^{k}}{2},\quad \Dten^{k}=\Uten^{k}-\Lten^{k}.
  \end{equation}
\end{itemize}
Note that a hyperprior entropy model initializes $(\Mten^{0}, \Sten^{0})$, while $(\Lten^{0}, \Uten^{0})$, and equivalently $(\Cten^{0}, \Dten^{0})$, define the initial feasible support of the latent elements.

At step $k$, the current feasible interval of $y_i$ is given by
$\mathcal{I}_i^{k-1}=[\Lten_i^{k-1},\,\Uten_i^{k-1}]$.
This interval is equally divided into three sub-intervals
$\mathcal{I}_{i,t}^{k-1}$ for $t=0,1,2$, corresponding to the next
undetermined trit at location $i$. BLI takes the current bound tensors
$(\Cten^{k-1},\Dten^{k-1})$ and the hyperprior parameters
$(\Mten^0,\Sten^0)$ as inputs, and estimates the Gaussian parameters
$(\Mten^k,\Sten^k)$ for entropy coding at step $k$. If $y_i$ lies in the
$t_i$-th sub-interval, the corresponding trit $t_i$ is encoded by an
arithmetic coder using
\begin{equation}
\textstyle
- \log_2 P(y_i \in \mathcal{I}_{i,t_i}^{k-1} \,|\, \Mten_i^k,\Sten_i^k)
\end{equation}
bits. Accordingly, BLI is trained to estimate $(\Mten^k,\Sten^k)$ by minimizing the rate loss
\begin{equation}
\textstyle
\mathcal{L}_{\mathit{R}}^{k}
=
- \sum_i \log_2 P(y_i \in \mathcal{I}_{i,t_i}^{k-1} \,|\, \Mten_i^k, \Sten_i^k).
\label{eq:rate_loss}
\end{equation}

In addition to the Gaussian parameters, BLI estimates a latent reconstruction within the current bounds by predicting a residual correction to the interval center:
\begin{equation}
\hat{\Yten}^{k-1} = \Cten^{k-1} + \Delta \hat{\Yten}^{k-1}.
\end{equation}
Here, the indices reflect the timing of the coding step; at step $k$, the Gaussian parameters $(\Mten^k,\Sten^k)$ are used to encode or decode the next subset of trits, whereas $\hat{\Yten}^{k-1}$ is the latent estimate consistent with the bounds $(\Cten^{k-1},\Dten^{k-1})$ available prior to coding that subset. After the subset is coded and the bounds are updated, the corresponding $\hat{\Yten}^{k}$ is estimated and used for reconstruction. The Gaussian parameters and latent estimates are produced from shared attention features but through separate prediction heads. The distortion loss is then defined as
\begin{equation}
\textstyle
\mathcal{L}_{\mathit{D}}^{k}
=
\| \Xten - \hat{\Xten}^{k} \|_2^2
=
\| \Xten - g_s(\hat{\Yten}^{k}) \|_2^2.
\label{eq:distortion_loss}
\end{equation}
The subset of trits is coded using $(\Mten^{k},\Sten^{k})$, and the decoded trits update the bounds from $(\Cten^{k-1},\Dten^{k-1})$ to $(\Cten^{k},\Dten^{k})$.

\begin{figure*}[t]
\centering
\includegraphics[width=\linewidth]{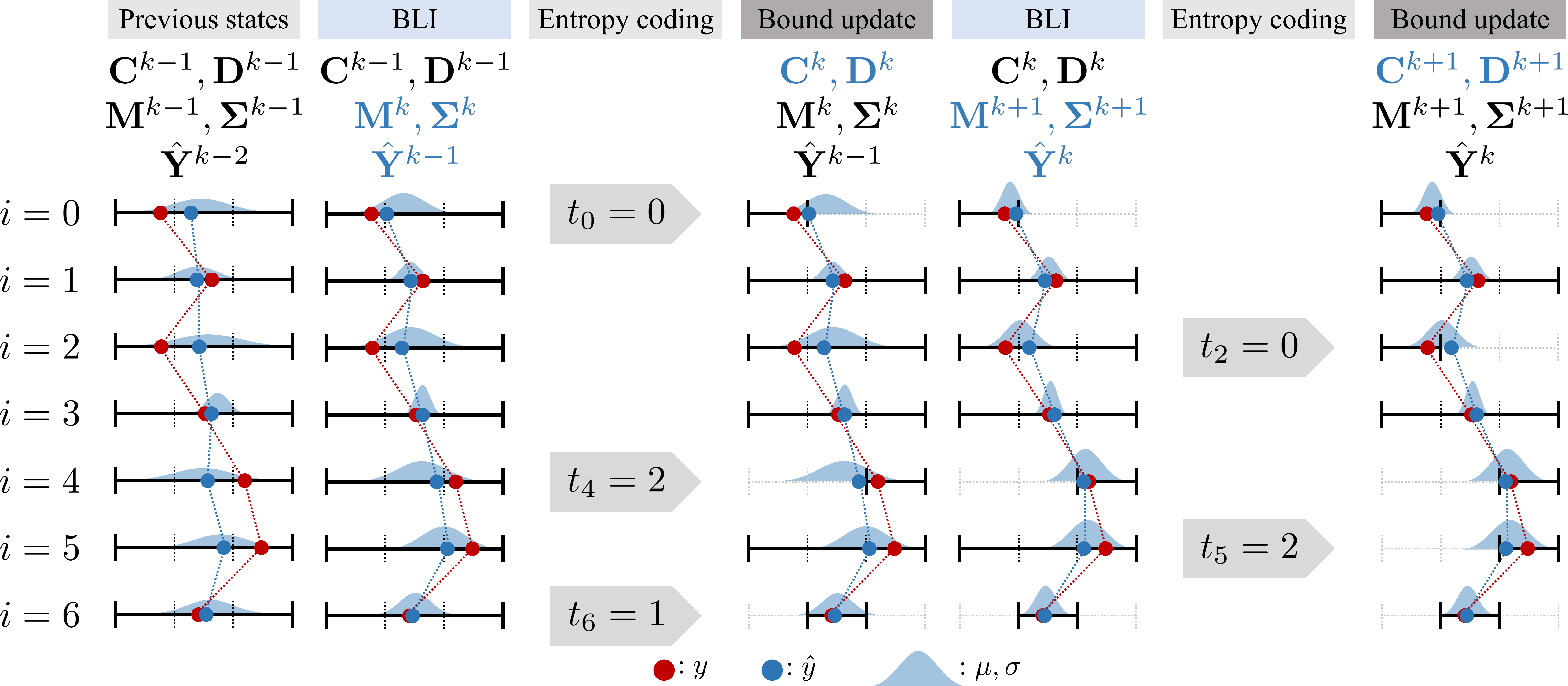}
\caption{
Illustration of the BLI refinement process using a 1D sequence of latent elements. Given the current bounds $(\Cten^{k-1},\Dten^{k-1})$, BLI estimates $(\Mten^k,\Sten^k)$ for entropy coding and $\hat{\Yten}^{k-1}$ for reconstruction under those bounds. After selected trits are decoded, the bounds are updated to $(\Cten^k,\Dten^k)$, yielding tighter feasible intervals for the next BLI step. Red dots denote original latent values, blue dots denote latent estimates, and blue densities denote estimated Gaussian distributions.
}
\label{fig:bli_illustration}
\end{figure*}

Figure~\ref{fig:bli_illustration} illustrates the refinement process using a 1D sequence of latent elements. At step $k$, the decoded trits define interval constraints $\mathcal{I}_i^{k-1}=[\Lten_i^{k-1},\Uten_i^{k-1}]$ for each latent element, and BLI performs learned constrained inference within these intervals. As additional trits are decoded, the intervals shrink monotonically, yielding tighter constraints for the next inference step. This process is conceptually related to denoising or inpainting under range constraints and to projection-based feasibility methods such as POCS \citep{y1996_SIAM_Bauschke}, since each step refines the estimates while enforcing consistency with known constraints. Unlike exact metric projection, however, BLI uses a learned amortized estimator trained for entropy modeling and reconstruction quality.

\subsection{BLI Codec Architecture}
\label{ssec:architecture}

Unlike previous trit-plane codecs \citep{y2022_CVPR_lee,y2023_CVPR_jeon} that reuse the initial Gaussian parameters $(\Mten^0,\Sten^0)$ throughout progressive decoding, BLI performs step-wise autoregressive refinement. Each trit-plane is divided into multiple coding steps, and both the Gaussian parameters $(\Mten^k,\Sten^k)$ and the latent estimate $\hat{\Yten}^{k-1}$ are inferred within the current feasible bounds at step $k$.

\begin{figure*}[t]
\centering
\includegraphics[width=\linewidth]{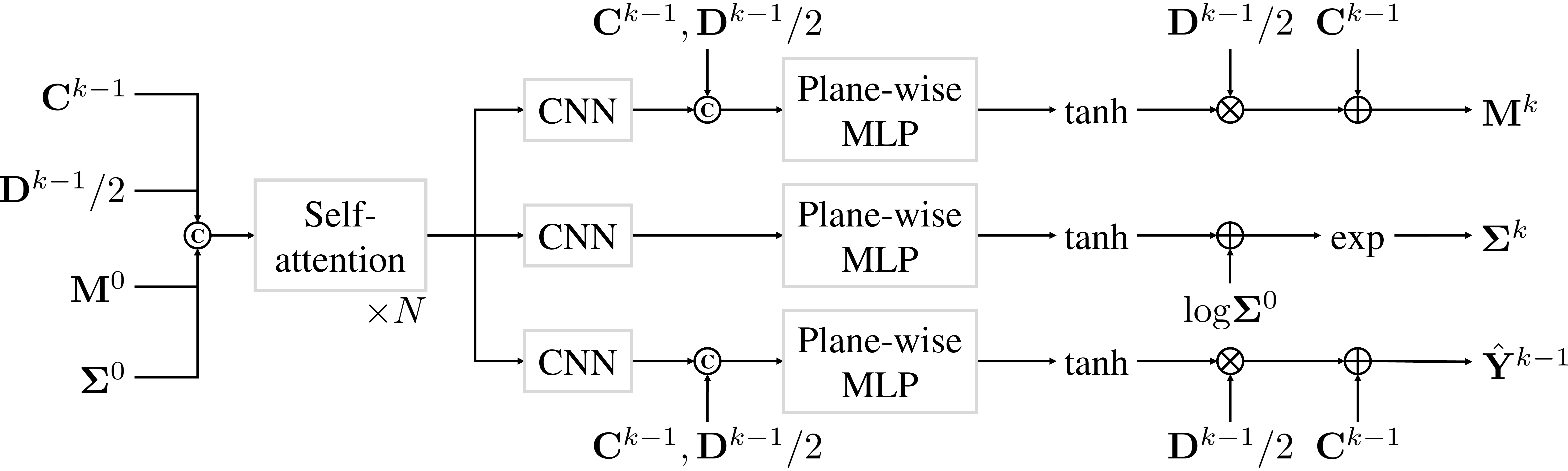}
\setlength{\abovecaptionskip}{-0.2cm}
\caption{
Architecture of the BLI module. BLI predicts the Gaussian parameters $(\Mten^k,\Sten^k)$ and the latent estimate $\hat{\Yten}^{k-1}$ from the current bounds and hyperprior parameters.
}
\label{fig:module_structures}
\end{figure*}

Before entropy coding at step $k$, the BLI module operates within the feasible interval $\mathcal{I}^{k-1}=[\Lten^{k-1},\Uten^{k-1}]$ determined by the trits decoded so far. As shown in Figure~\ref{fig:module_structures}, the BLI module receives the interval center $\Cten^{k-1}$, the half-width $\Dten^{k-1}/2$, and the hyperprior parameters $(\Mten^0,\Sten^0)$ as inputs. A stack of Swin attention blocks \citep{y2021_ICCV_liu} extracts bound-conditioned features by modeling spatial dependencies among partially decoded latent elements. These features are then processed by CNN blocks and plane-wise MLP heads to estimate $(\Mten^k,\Sten^k)$ and $\hat{\Yten}^{k-1}$. The plane-wise MLP heads allow the module to adapt its prediction to each trit-plane level.

For the estimation of $\Mten^k$ and $\hat{\Yten}^{k-1}$, the CNN features are additionally concatenated with $\Cten^{k-1}$ and $\Dten^{k-1}/2$, since the interval center provides a natural anchor and the half-width specifies the allowable residual range. The outputs for $\Mten^k$ and $\hat{\Yten}^{k-1}$ are processed through a $\tanh$ activation and rescaled by $\Dten^{k-1}/2$ around $\Cten^{k-1}$, ensuring consistency with the feasible interval $\mathcal{I}^{k-1}$. As illustrated in Figures~\ref{fig:overview} and \ref{fig:bli_illustration}, the estimates at step $k$ are used to code the next subset of trits, after which the bounds are updated to $(\Cten^k,\Dten^k)$. As the decoded trits progressively tighten the feasible bounds, this step-wise refinement improves the accuracy of entropy parameter estimation and reconstruction.

\subsection{Implementation}
\label{ssec:implementation}

We embed the proposed BLI framework into FTIC \citep{y2024_ICLR_li}, a state-of-the-art learned image codec. The original entropy model in FTIC performs channel-wise autoregressive prediction for standard fixed-rate decoding. This design is not directly compatible with trit-plane coding, where each latent element may be only partially decoded at intermediate steps. We therefore replace the channel-wise autoregressive entropy model with a non-autoregressive hyperprior variant that directly estimates the initial Gaussian parameters $(\Mten^0,\Sten^0)$ from the hyper latent tensor $\Zten$.

Entropy coding is performed using the ANS algorithm \citep{y2013_arXiv_duda_ans}. We train the codecs on the Flickr2W dataset \citep{y2020_arXiv_liu_flickr} using $256\times256$ patches, a batch size of 4, and the AdamW optimizer \citep{y2019_ICLR_Loshchilov}. Training is conducted on an NVIDIA RTX A6000 GPU.

The BLI codec is optimized in two stages. First, we train the modified FTIC backbone at the highest bitrate point. It is initialized from pretrained FTIC weights and optimized for $1.0$M iterations with a learning rate of $10^{-4}$ using the corresponding Lagrangian multiplier. This trained backbone is used as the common basis for BLI and for the DPICT \citep{y2022_CVPR_lee} and CTC \citep{y2023_CVPR_jeon} baselines, with the additional training scheme of \citet{y2023_CVPR_jeon} applied to CTC.

Second, we optimize the BLI codec using the following combined loss:
\begin{equation}
\label{eq:loss_total}
\mathcal{L}
=
\sum_{l\in\mathcal{K}_{\mathrm{end}}} \alpha_l
\Bigl(
\sum_{k=1}^{l}\mathcal{L}_{\mathit{R}}^{k}
+
\lambda \mathcal{L}_{\mathit{D}}^{l}
\Bigr),
\end{equation}
where $\mathcal{K}_{\mathrm{end}}$ denotes the set of endpoint step indices of the trit-planes. $\mathcal{L}_{\mathit{R}}^k$ and $\mathcal{L}_{\mathit{D}}^l$ are defined in \eqref{eq:rate_loss} and \eqref{eq:distortion_loss}, respectively. The rate term is accumulated up to each endpoint step $l\in\mathcal{K}_{\mathrm{end}}$, so that all coding steps up to the corresponding trit-plane endpoint contribute to the loss. The weight $\alpha_l$ balances the contributions of endpoint losses at different trit-plane levels, because the corresponding quantization intervals have different scales; its detailed setting is provided in Appendix~\ref{ssec:training_loss}. We set $\lambda = 0.0483$ and optimize BLI for $0.5$M iterations with a learning rate of $5\times10^{-5}$.

For transmission scheduling, we use random trit assignments during training and the R-D prioritized transmission order of \citet{y2022_CVPR_lee} during evaluation. During training, each trit-plane is divided into four steps, \ie $\mathcal{K}_{\mathrm{end}}=\{4,8,12,16,20\}$, with trits randomly assigned across steps to expose BLI to diverse partial-observation patterns. During evaluation, we use $\mathcal{K}_{\mathrm{end}}=\{4,8,14,30,54\}$, since the trit-planes closer to the LST typically contain more high-entropy trits and therefore benefit more from finer subdivision. Further implementation details are provided in Appendix~\ref{sec:impl_details}.

\section{Experimental Results}
\label{sec:experiment}

We evaluate the proposed BLI codec on the Kodak Lossless dataset~\citep{kodak}, the CLIC 2021 validation set~\citep{clic2021}, and the JPEG-AI test set~\citep{jpegai}. Kodak contains 24 images at resolutions of $512\times768$ or $768\times512$, while CLIC and JPEG-AI contain 41 and 16 higher-resolution images, respectively. Compression performance is measured in bits per pixel (bpp), PSNR, and MS-SSIM, and summarized using BD-rate~\citep{bdrate_excel}.
 
\subsection{Comparative Assessment}
\label{ssec:comparison}

\begin{figure*}[t]
\centering
\includegraphics[width=\linewidth]{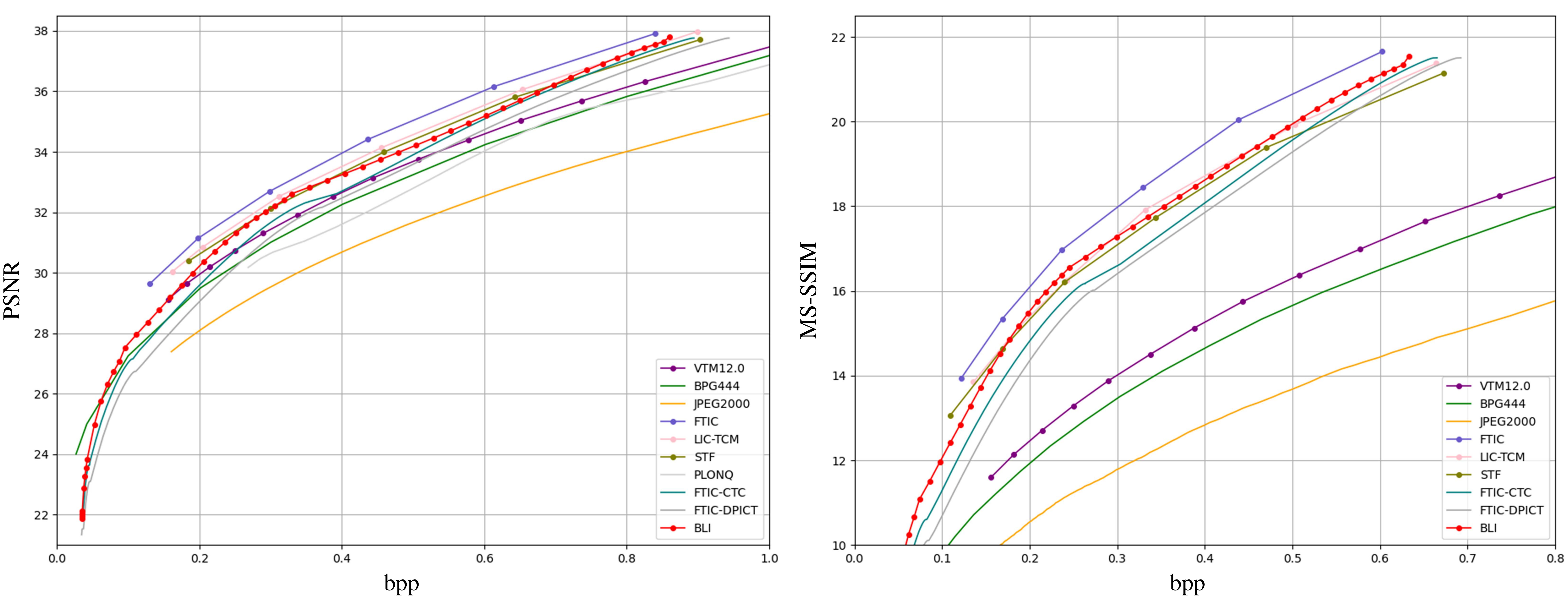}
\setlength{\abovecaptionskip}{-0.1cm}
\caption{
R-D curves on Kodak against traditional, fixed-rate learned, and progressive learned codecs. The left plot reports PSNR, and the right plot reports MS-SSIM in dB.
}
\label{fig:RDcurves}
\end{figure*}

\noindent
\textbf{R-D curve comparison:}
We compare BLI with representative image compression methods: JPEG2000 \citep{y2001_SPM_JPEG2000}, BPG444 \citep{bpg}, and VTM 12.0 \citep{y2021_TCSVT_Bross} as traditional codecs; STF \citep{y2022_CVPR_zou}, LIC-TCM \citep{y2023_CVPR_liu}, and FTIC \citep{y2024_ICLR_li} as fixed-rate learned codecs; and PLONQ \citep{y2021_ICIP_lu}, DPICT \citep{y2022_CVPR_lee}, and CTC \citep{y2023_CVPR_jeon} as progressive codecs. Since earlier progressive methods were built on CNN-based baselines~\citep{y2018_NIPS_minnen,y2020_CVPR_cheng}, we re-implement DPICT and CTC on the stronger FTIC backbone for fair comparison, denoted as FTIC-DPICT and FTIC-CTC. Figure~\ref{fig:RDcurves} shows the PSNR and MS-SSIM R-D curves on the Kodak dataset, with MS-SSIM converted to dB. BLI consistently outperforms existing progressive codecs over a wide bitrate range, with particularly strong gains at low-to-medium bitrates, where limited decoded information makes bound-conditioned inference especially important for reducing reconstruction distortions. Additional R-D curves and BD-rate results on the CLIC and JPEG-AI datasets are provided in Appendix~\ref{sec:appendix_exp}.

\begin{table}[t]
\centering
\begin{minipage}[t]{0.49\linewidth}
\centering
\caption{PSNR BD-rates (\%) with respect to FTIC-DPICT.}
\label{table:bdrates}
\vspace{0.21cm}
\resizebox{\linewidth}{!}{
    \begin{tabular}{@{} l | l | r r r@{} }
        \toprule
        \multicolumn{2}{c}{} & Kodak & CLIC & JPEG-AI \\
        \midrule

        \multirow{3}{*}{Traditional} & JPEG2000~\citep{y2001_SPM_JPEG2000} & $47.54$ & $58.41$ & $55.73$ \\
        & BPG444~\citep{bpg} & $-7.92$ & $0.21$ & $0.74$ \\
        & VTM 12.0~\citep{y2021_TCSVT_Bross} & $-2.14$ & $-2.17$ & $-0.85$ \\

        \midrule

        \multirow{3}{*}{Fixed-rate} & STF~\citep{y2022_CVPR_zou} & $-14.85$ & $-12.87$ & $-11.69$ \\
        & LIC-TCM~\citep{y2023_CVPR_liu} & $-18.69$ & $-19.37$ & $-18.18$ \\
        & FTIC~\citep{y2024_ICLR_li} & $-26.37$ & $-23.33$ & $-20.49$ \\

        \midrule

        \multirow{3}{*}{Progressive} & PLONQ~\citep{y2021_ICIP_lu} & $14.83$ & - & $21.29$ \\
        & FTIC-CTC~\citep{y2023_CVPR_jeon} & $-9.82$ & $-11.52$ & $-11.53$ \\
        & BLI & $-17.78$ & $-17.12$ & $-17.52$ \\
        \bottomrule
    \end{tabular}
}
\end{minipage}\hfill
\begin{minipage}[t]{0.49\linewidth}
\centering
\caption{
Comparison of parameter counts and latencies. The channel-wise autoregressive entropy model of FTIC is removed for progressive methods, and the DPICT postprocessing network is not used.
}
\label{table:complexities}
\resizebox{\linewidth}{!}{
    \begin{tabular}{@{} l | l | c | c @{} }
        \toprule
        \multicolumn{2}{c}{} & \# Params. (\text{M}) & Latency (\text{s}) \\

        \midrule
        
        \multirow{2}{*}{Fixed-rate} & LIC-TCM~\citep{y2023_CVPR_liu} & $76.57$ & $6.53$ \\
        & FTIC~\citep{y2024_ICLR_li} & $71.01$ & $215.73$ \\

        \midrule

        \multirow{3}{*}{Progressive} & FTIC-DPICT~\citep{y2022_CVPR_lee} & $40.53$ & $5.85$ \\
        & FTIC-CTC~\citep{y2023_CVPR_jeon} & $1,032.65$ & $6.62$ \\
        & BLI & $196.12$ & $4.77$ \\
        \bottomrule
    \end{tabular}
}
\end{minipage}
\end{table}

Compared with fixed-rate learned codecs, BLI remains competitive despite the additional constraint of progressive decoding. In particular, BLI approaches the performance of strong fixed-rate methods and surpasses STF over most of the MS-SSIM range. This suggests that bound-conditioned refinement helps preserve structural and perceptual details. Note that STF, LIC-TCM, and FTIC are individually optimized for each target rate, whereas BLI generates all R-D points from a single progressive codec.

Among progressive approaches, BLI achieves the best R-D performance over most of the bitrate range. Its consistent advantage over FTIC-DPICT reflects the benefit of bound-conditioned inference. The gains over FTIC-CTC arise from finer step-wise autoregressive modeling within each trit-plane, which enables more effective use of spatial context and bound information for entropy modeling and latent reconstruction.

\begin{figure*}[t]
\centering
\includegraphics[width=\linewidth]{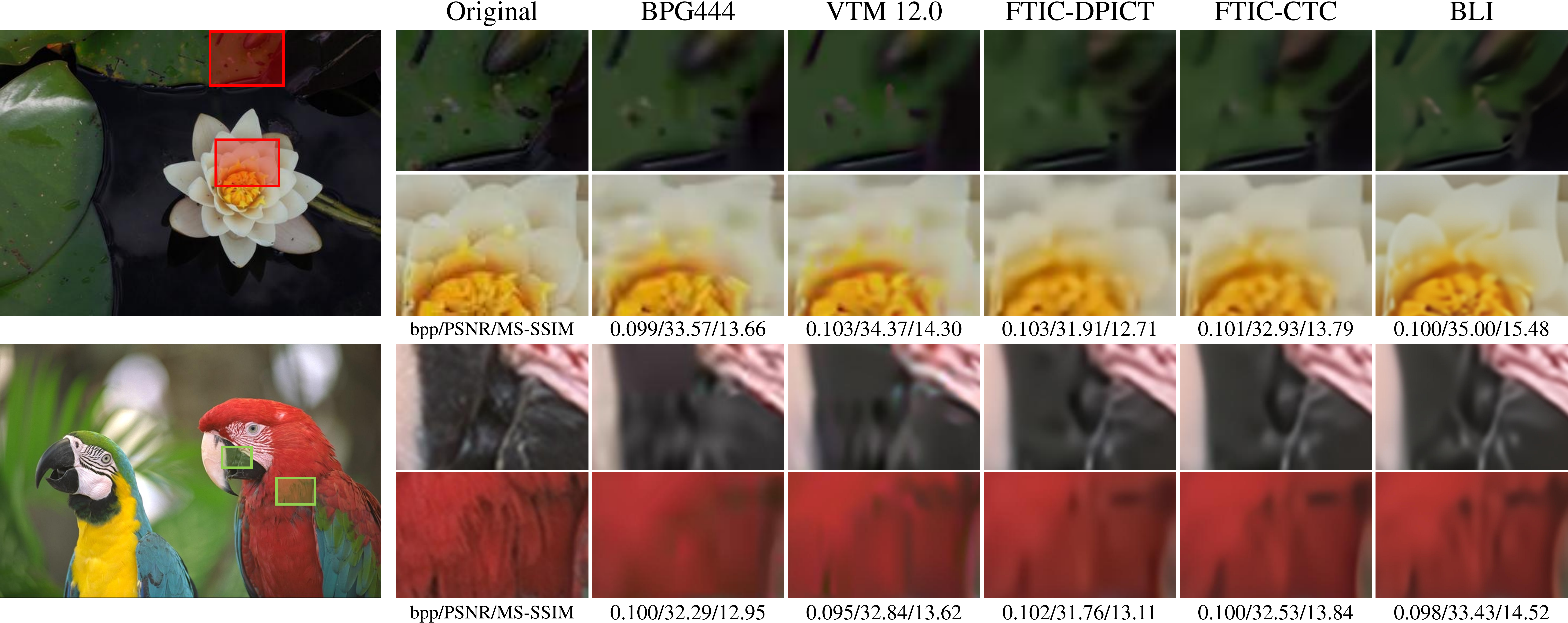}
\setlength{\abovecaptionskip}{-0.2cm}
\caption{Qualitative comparison against representative traditional and progressive codecs.}
\label{fig:qualitative}
\end{figure*}

\begin{figure*}[t]
\centering
\includegraphics[width=\linewidth]{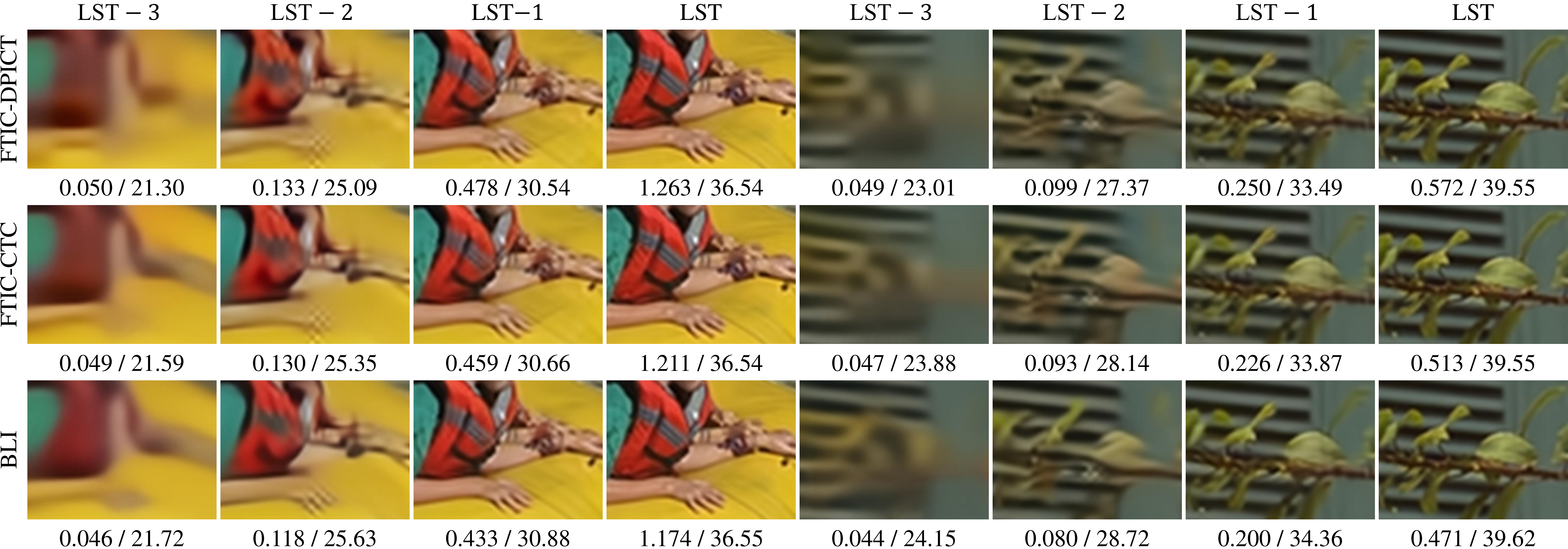}
\caption{
Qualitative comparison with existing trit-plane codecs at trit-plane endpoints. Within each column, the same number of trits is transmitted for all three codecs. The numbers below each result indicate bpp and PSNR.
}
\label{fig:qual_endpoints}
\end{figure*}

\textbf{BD-rate analysis:}
Table~\ref{table:bdrates} reports PSNR BD-rates relative to FTIC-DPICT. BLI reduces bitrate by $17.78\%$, $17.12\%$, and $17.52\%$ on Kodak, CLIC, and JPEG-AI, respectively, showing consistent gains within the progressive setting. These improvements stem from flexible in-plane autoregression, which exploits spatial context and bound information at multiple coding steps. In contrast, FTIC-CTC does not perform step-wise bound-conditioned refinement within each trit-plane, limiting its coding efficiency. Overall, BLI provides a clear R-D advantage for progressive image compression.

\noindent
\textbf{Model complexity:}
Table~\ref{table:complexities} compares the parameter counts and latency of BLI with fixed-rate and progressive baselines. BLI has about $5.3\times$ fewer parameters than FTIC-CTC, which uses a separate prediction network for each trit-plane and therefore incurs substantial parameter overhead. Although BLI requires more parameters than FTIC-DPICT, it replaces plane-specific predictors with a shared BLI module and achieves the lowest latency among the compared codecs. This latency reduction stems from efficient Gaussian interval integration, which lowers non-neural computation compared with the loop-based discrete probability composition used in FTIC-DPICT and FTIC-CTC.

\noindent
\textbf{Qualitative results:}
Figure~\ref{fig:qualitative} compares reconstructed images at nearly identical bitrates. Traditional codecs such as BPG444 and VTM 12.0 introduce noticeable blurring around petal boundaries and feather textures. FTIC-DPICT also suffers from blurred details, while FTIC-CTC produces sharper structures but still fails to recover fine textures faithfully. In contrast, BLI better preserves high-frequency details and produces fewer noticeable artifacts overall. This improvement comes from BLI's bound-conditioned autoregressive refinement, which uses spatial context and interval constraints to recover fine structures from partially decoded latents.

Figure~\ref{fig:qual_endpoints} compares BLI with existing trit-plane codecs at the endpoints of trit-planes, where the same number of trits is transmitted within each column. BLI consistently achieves the best R-D performance, followed by FTIC-CTC and FTIC-DPICT. This shows that, even at the same trit-plane endpoint, BLI encodes the transmitted trits with fewer bits while producing more accurate latent estimates. The results further confirm that bound-conditioned inference improves both entropy modeling and reconstruction at intermediate progressive decoding stages.

\subsection{Analysis}
\label{ssec:analysis}

\begin{figure}[t]
\centering
\begin{tabular}{@{}p{0.48\textwidth}@{\hspace{0.04\textwidth}}p{0.48\textwidth}@{}}
\begin{minipage}[t]{\linewidth}
\vspace{0.1cm}
\captionsetup{type=table,aboveskip=0pt,belowskip=2mm}
\captionof{table}{
Comparison of ablation variants and per-plane refinement schedules on the Kodak dataset. BD-rates are measured relative to FTIC-DPICT.
}
\label{table:ablation}
\vspace{0.3cm}
\resizebox{\linewidth}{!}{
\begin{tabular}{@{} l | c c c@{} }
    \toprule
     & \# Params. (\text{M}) & Latency (\text{s}) & BD-rate (\%) \\
    \midrule
    Method \Romannum{1} & $155.28$ & $4.74$ & $-15.64$ \\
    Method \Romannum{2} & $139.41$ & $4.58$ & $-15.01$ \\
    Method \Romannum{3} & $119.32$ & $4.00$ & $-12.48$ \\
    \midrule
    1 step/plane & $196.12$ & $3.82$ & $-10.40$ \\
    2 steps/plane & $196.12$ & $3.92$ & $-12.90$ \\
    4 steps/plane & $196.12$ & $4.15$ & $-14.79$ \\
    \midrule
    BLI & $196.12$ & $4.77$ & $-17.78$ \\
    \bottomrule
\end{tabular}}
\end{minipage}
&
\begin{minipage}[t]{\linewidth}
\vspace{0pt}
\centering
\includegraphics[width=\linewidth]{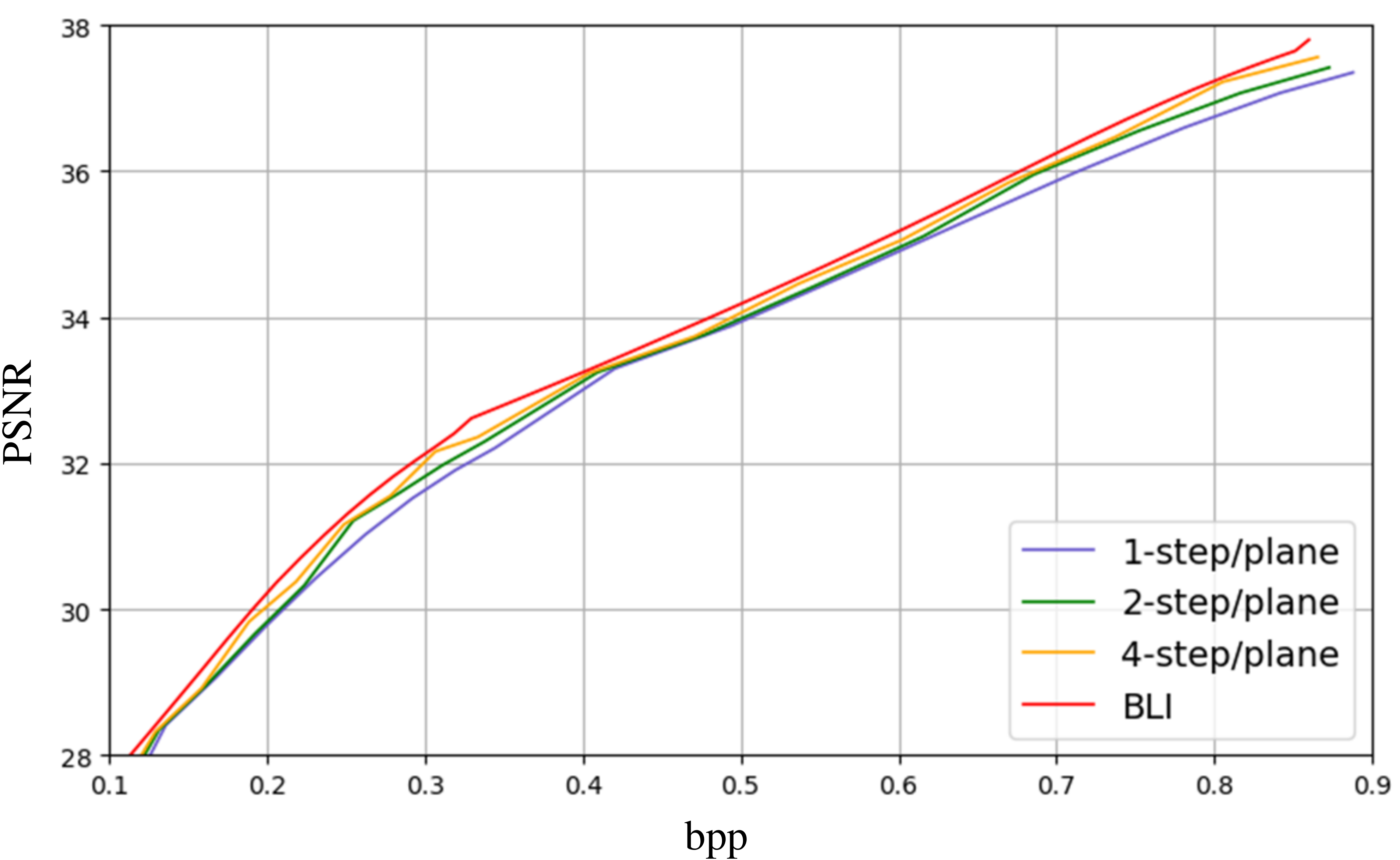}
\caption{Effect of the number of refinement steps per trit-plane on the Kodak dataset.}
\label{fig:steps_analysis}
\end{minipage}
\end{tabular}
\vspace*{-0.3cm}
\end{figure}

\noindent
\textbf{Model design ablations:}
We evaluate three variants to examine the effects of the bound-related model design, the reconstruction prediction branch, and module capacity. Table~\ref{table:ablation} summarizes the results. In method \Romannum{1}, we remove $\Dten^{k-1}/2$ from the BLI input and disable the interval-rescaled output constraint for $\Mten^k$ and $\hat{\Yten}^{k-1}$. In method \Romannum{2}, we remove the reconstruction prediction branch and instead use the expectation computed from the predicted Gaussian distribution for reconstruction. In method \Romannum{3}, we reduce the hidden dimension of the attention module and the number of CNN layers.

As shown in Table~\ref{table:ablation}, method \Romannum{1} results in a BD-rate of $-15.64\%$, indicating that the bound width and interval-consistent prediction are important for effective progressive inference. Method \Romannum{2} yields $-15.01\%$, showing that directly estimating the latent reconstruction is more effective than relying on the Gaussian expectation. Method \Romannum{3} gives $-12.48\%$, confirming the effect of module capacity.

\begin{figure*}[t]
\centering
\includegraphics[width=\linewidth]{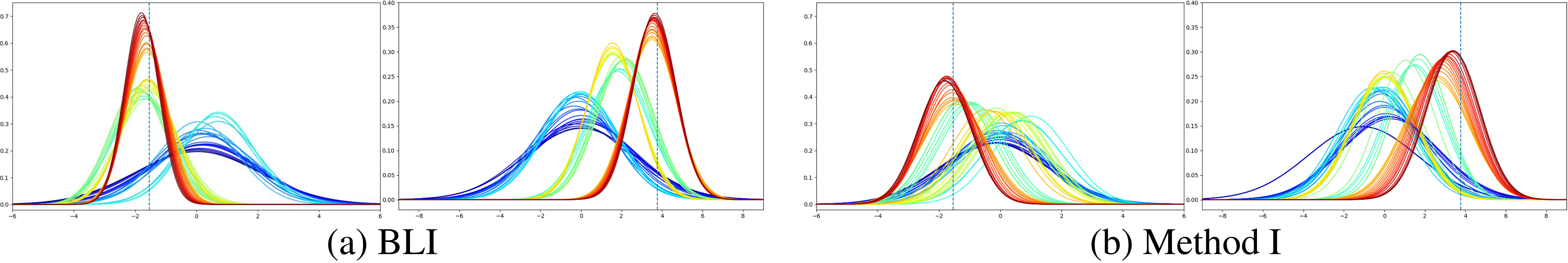}
\setlength{\abovecaptionskip}{-0.2cm}
\caption{
Step-wise updates to predicted Gaussian distributions for two latent elements during evaluation. We compare BLI with method \Romannum{1}, the ablation variant without the bound-width input and interval-rescaled output constraint. Colors transition from blue to red as decoding proceeds.
}
\label{fig:visualize_dist}
\end{figure*}

\noindent
\textbf{Step-wise distribution refinement:}
Figure~\ref{fig:visualize_dist} illustrates how the predicted Gaussian distributions for selected latent elements evolve during evaluation. Each curve corresponds to the predicted distribution at a particular step, with colors transitioning from blue (early steps) to red (later steps). The vertical dashed line indicates the ground-truth latent value. Compared with method \Romannum{1}, BLI shifts the predicted distributions more consistently toward the ground-truth latent and shows clearer variance reduction over successive steps. This confirms that the bound-width input and interval-rescaled output constraint help refine uncertainty during progressive decoding.

\noindent
\textbf{Refinement steps per trit-plane:}
Table~\ref{table:ablation} and Figure~\ref{fig:steps_analysis} analyze the effect of the number of refinement steps within each trit-plane on the Kodak dataset. We compare schedules using one, two, or four steps per trit-plane with the full BLI schedule, \ie the evaluation schedule with $\mathcal{K}_{\mathrm{end}}=\{4,8,14,30,54\}$. As the number of steps increases, R-D performance improves consistently: the BD-rate decreases from $-10.40\%$ for the 1-step setting to $-12.90\%$ and $-14.79\%$ for the 2-step and 4-step settings, respectively, and further reaches $-17.78\%$ for the full setting. Notably, even the 1-step configuration slightly outperforms FTIC-CTC ($-10.40\%$ vs. $-9.82\%$), despite using far fewer parameters and lower latency. These results show that BLI achieves strong R-D performance through more frequent in-plane autoregressive refinement, which allows bound information and spatial context to be used more effectively throughout progressive decoding.

\noindent
\textbf{Cross-backbone generalization:} We further evaluate BLI on two backbones of learned image codecs beyond FTIC: the CNN-based backbone of \citet{y2020_CVPR_cheng}, used in the original DPICT and CTC papers, and \citet{y2023_CVPR_liu}. Table~\ref{table:cross_backbone} reports BD-rates with respect to the corresponding DPICT baseline. On the backbone of \citet{y2020_CVPR_cheng}, BLI achieves $-19.87\%$, compared with the $-14.84\%$ reported for CTC in its original paper. On the backbone of \citet{y2023_CVPR_liu}, we train CTC and BLI under the same training configurations. BLI improves the BD-rate from $-6.86\%$ to $-22.30\%$. These results indicate that the benefit of BLI is preserved across different $g_a$ and $g_s$ backbones.

\noindent
\textbf{Discrete vs. continuous distributions:} To distinguish the effect of BLI's continuous distribution modeling from that of in-plane autoregressive refinement, we replace Gaussian interval integration with direct ternary-logit prediction, similar to CTC, while retaining the same context module, reconstruction branch and refinement schedule. Table~\ref{table:discrete_control} compares this variant, denoted ``IPAR + discrete'', with BLI at matched refinement depths. BLI achieves lower BD-rates with one step per plane ($-10.40\%$ vs. $-9.30\%$) and with the full schedule ($-17.78\%$ vs. $-14.06\%$), despite using fewer parameters. Thus, increasing the number of refinement steps benefits both methods, while direct discrete prediction does not reproduce the gains of continuous interval-based distribution modeling.

\subsection{Limitations and Future Work}
\label{ssec:limitations}

Although BLI supports flexible transmission schedules within each trit-plane, our current evaluation adopts the R-D-prioritized transmission order of \citet{y2022_CVPR_lee}. Designing transmission orders better matched to bound-conditioned inference remains an important direction for future work. A performance gap also remains between BLI and recent fixed-rate codecs, as expected for progressive codecs that reconstruct from bitstream prefixes. Closing this gap is another important direction for future work.

\begin{table}[t]
\centering
\begin{minipage}[t]{0.43\linewidth}
\centering
\caption{
Cross-backbone PSNR BD-rates (\%) with respect to the corresponding DPICT baseline.
}
\label{table:cross_backbone}
\vspace{0.1cm}
\resizebox{0.88\linewidth}{!}{
    \begin{tabular}{@{}l|rr@{}}
        \toprule
        Backbone & CTC & BLI \\
        \midrule
        \citet{y2020_CVPR_cheng} & $-14.84$ & $-19.87$ \\
        \citet{y2023_CVPR_liu} & $-6.86$ & $-22.30$ \\
        \bottomrule
    \end{tabular}
}

\smallskip

\end{minipage}\hfill
\begin{minipage}[t]{0.53\linewidth}
\centering
\caption{Controlled comparison with discrete prediction. PSNR BD-rates (\%) are relative to FTIC-DPICT.}
\label{table:discrete_control}
\vspace{0.45cm}
\resizebox{\linewidth}{!}{%
    \begin{tabular}{@{}l|r|rr@{}}
        \toprule
        Method & Params. (M) & 1 step/plane & Full \\
        \midrule
        IPAR + discrete & $207.80$ & $-9.30$ & $-14.06$ \\
        BLI             & $196.12$ & $-10.40$ & $-17.78$ \\
        \bottomrule
    \end{tabular}%
}
\end{minipage}
\vspace*{-0.2cm}
\end{table}

\section{Conclusions}
\label{sec:conclusions}
We introduced bound-conditioned latent inference (BLI), a progressive image compression framework that reformulates trit-plane coding as sequential latent inference under shrinking interval constraints. By jointly refining the Gaussian parameters for entropy modeling and the latent estimates for reconstruction within progressively updated bounds, BLI enables flexible in-plane autoregressive refinement and more effective use of the information shared by the encoder and decoder at each coding step. Experiments show that BLI achieves state-of-the-art R-D performance among progressive codecs and preserves high-frequency details more faithfully, while replacing CTC's plane-specific predictors with a shared BLI module that uses fewer parameters and achieves lower latency. Overall, BLI provides an effective framework for progressive learned image compression.

\begin{ack}
This work was supported by the National Research Foundation of Korea (NRF) grants funded by the Korea government(MSIT) (No. RS-2024-00397293 and No. RS-2026-25616634).

The authors declare no competing interests.

\end{ack}

\bibliographystyle{plainnat}
\bibliography{main}


\clearpage
\appendix

\section{Additional Experimental Results}
\label{sec:appendix_exp}

\begin{figure}[!htbp]
\centering
\includegraphics[width=\linewidth]{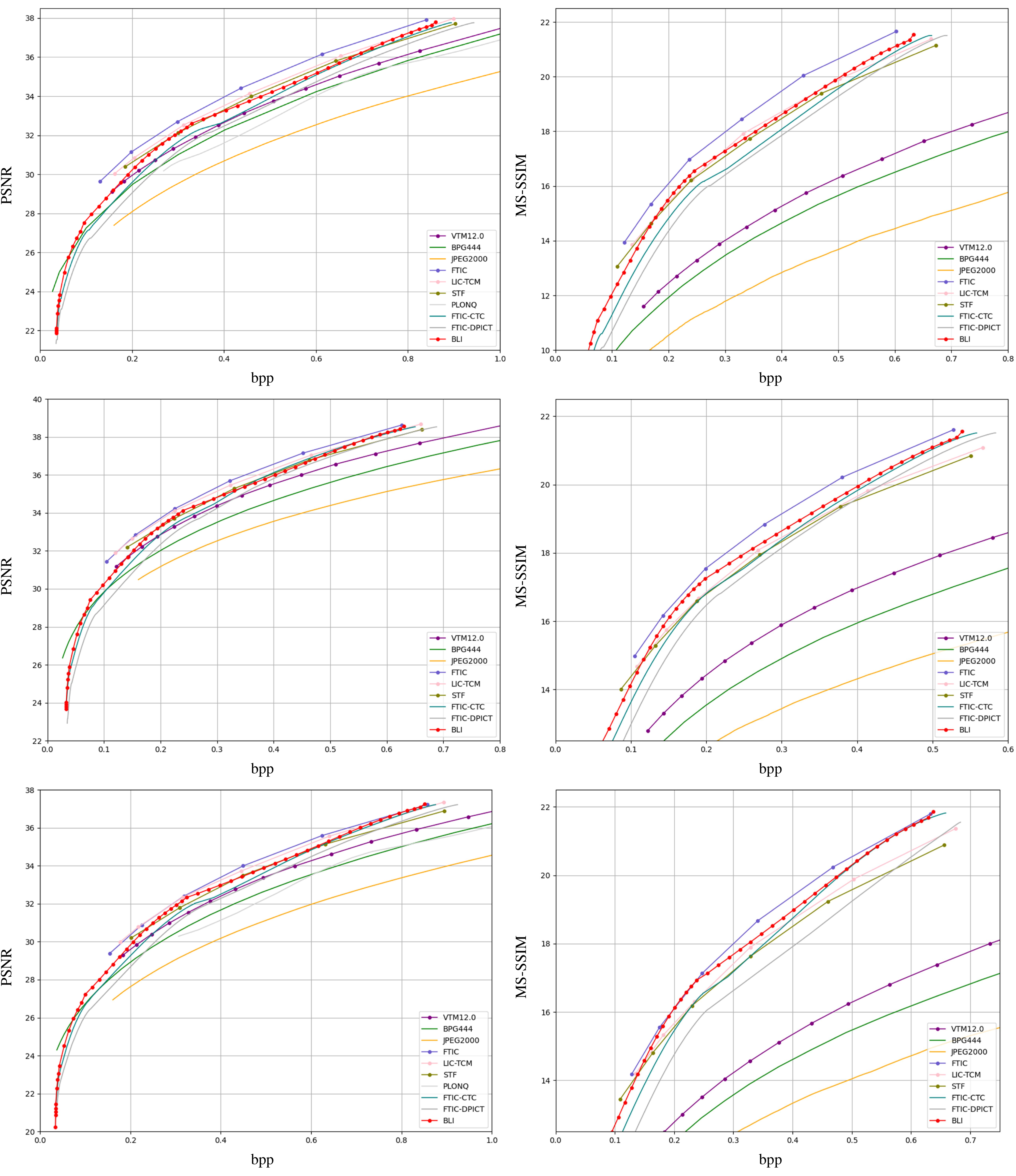}
\setlength{\abovecaptionskip}{-0.2cm}
\caption{R-D curve comparison of the proposed BLI codec with existing progressive and fixed-rate image codecs. From top to bottom, the rows correspond to Kodak, CLIC 2021, and JPEG-AI. Best viewed digitally with zooming.}
\label{fig:appendix_rdcurves}
\vspace*{-0.7cm}
\end{figure}

\textbf{More R-D curves:} 
For completeness, we provide additional comparative results on the Kodak, CLIC 2021, and JPEG-AI datasets. Figure~\ref{fig:appendix_rdcurves} presents the full PSNR and MS-SSIM R-D curves for all three datasets. Table~\ref{table:appendix_bdrates} summarizes the BD-rates for both PSNR and MS-SSIM with respect to FTIC-DPICT.

\begin{table}[t]
\centering
\caption{BD-rates (\%) with respect to FTIC-DPICT.}
\label{table:appendix_bdrates}
\resizebox{\linewidth}{!}{
    \begin{tabular}{@{} l | l | r r r | r r r@{} }
        \toprule
        \multicolumn{2}{c}{} & \multicolumn{3}{c|}{PSNR} & \multicolumn{3}{c}{MS-SSIM} \\
        \cmidrule{3-8}
        
        \multicolumn{2}{c}{} & Kodak & CLIC & JPEG-AI & Kodak & CLIC & JPEG-AI \\
        \midrule

        \multirow{3}{*}{Traditional} & JPEG2000~\citep{y2001_SPM_JPEG2000} & $47.54$ & $58.41$ & $55.73$ & $183.09$ & $232.17$ & $205.86$ \\
        & BPG444~\citep{bpg} & $-7.92$ & $0.21$ & $0.74$ & $64.25$ & $74.67$ & $89.49$ \\
        & VTM 12.0~\citep{y2021_TCSVT_Bross} & $-2.14$ & $-2.17$ & $-0.85$ & $69.49$ & $72.06$ & $94.68$ \\

        \midrule

        \multirow{3}{*}{Fixed-rate} & STF~\citep{y2022_CVPR_zou} & $-14.85$ & $-12.87$ & $-11.69$ & $-12.64$ & $-7.73$ & $-5.57$ \\
        & LIC-TCM~\citep{y2023_CVPR_liu} & $-18.69$ & $-19.37$ & $-18.18$ & $-13.84$ & $-7.74$ & $-6.53$ \\
        & FTIC~\citep{y2024_ICLR_li} & $-26.37$ & $-23.33$ & $-20.49$ & $-25.27$ & $-20.44$ & $-16.58$ \\

        \midrule

        \multirow{3}{*}{Progressive} & PLONQ~\citep{y2021_ICIP_lu} & $14.83$ & - & $21.29$ & - & - & - \\
        & FTIC-CTC~\citep{y2023_CVPR_jeon} & $-9.82$ & $-11.52$ & $-11.53$ & $-8.43$ & $-10.58$ & $-9.20$ \\
        & BLI & $-17.78$ & $-17.12$ & $-17.52$ & $-18.56$ & $-18.14$ & $-17.61$ \\
        \bottomrule
    \end{tabular}
}
\end{table}

\section{Implementation Details}
\label{sec:impl_details}

\subsection{Architecture Configurations}
\label{ssec:arch_configs}

We embed the proposed BLI framework into FTIC \citep{y2024_ICLR_li}, where the hyper latent $\mathbf{Z}$ has 128 channels and the primary latent $\mathbf{Y}$ has 320 channels. To estimate the initial Gaussian parameters $(\mathbf{M}^{0},\mathbf{\Sigma}^{0})$, we employ two identical predictor networks. Each predictor consists of a residual CNN block with three $3\times3$ convolution layers, followed by two FAT blocks \citep{y2024_ICLR_li}, and a final $3\times3$ convolution layer, yielding a 320-channel output tensor that provides the initial distribution estimates. All latent-related tensors, including $\Yten$, $\Mten^k$, $\Sten^k$, $\Lten^k$, $\Uten^k$, $\Cten^k$, and $\Dten^k$, have shape $C\times H\times W$, where $C=320$ and $(H,W)$ are $1/16$ of the input image resolution. In the BLI model architecture, a total of 10 Swin attention \citep{y2021_ICCV_liu} layers are used, each with a hidden dimension of 960, 16 attention heads, and a window size of 8. Each CNN branch uses four depthwise convolution \citep{y2017_CVPR_chollet} layers to improve efficiency while preserving spatial modeling capacity.

\subsection{Training Loss}
\label{ssec:training_loss}
In \eqref{eq:loss_total}, the endpoint indices are given by $\mathcal{K}_{\mathrm{end}}=\{4,8,12,16,20\}$, where $l=4$ and $l=20$ represent the MST and the LST, respectively. The weight $\alpha_l$ is adopted to account for the different quantization intervals across plane levels. Specifically, because the quantization interval increases by a factor of $3$ from the LST to the MST, we set $\alpha_l$ proportional to the inverse squared quantization interval. As a result, the weights decrease by a factor of $1/9$ from the LST to the MST, corresponding to $\{(1/9)^0, (1/9)^1, \ldots, (1/9)^4\}$ from the LST to the MST.

\subsection{Transmission Schedule}
\label{ssec:trans_schedule}

In BLI, the trits transmitted within each trit-plane are divided into multiple coding steps. The step endpoints are represented by $\mathcal{K}_{\mathrm{end}}$, which gives the cumulative coding-step indices at the endpoints of successive trit-planes. During training, each trit-plane is uniformly divided into four steps, yielding $\mathcal{K}_{\mathrm{end}}=\{4,8,12,16,20\}$. To expose BLI to diverse partial-observation patterns, we randomly assign trits within each trit-plane to the four steps during the initial training stage. We then fine-tune the codec using the same R-D-prioritized transmission order \citep{y2022_CVPR_lee} as in evaluation.

During evaluation, we use a nonuniform full BLI schedule, $\mathcal{K}_{\mathrm{end}}=\{4,8,14,30,54\}$, corresponding to step allocations of $(4,4,6,16,24)$ from the MST to the LST. This schedule allocates more refinement steps to trit-planes closer to the LST, since these planes typically contain more high-entropy trits and benefit more from finer in-plane subdivision. Within each trit-plane, the assignment of trits to coding steps follows the R-D-prioritized order. Specifically, for each latent element, we estimate the expected bit increase $\Delta R$ and the distortion change $\Delta D$ using the predicted distribution parameters available at the beginning of the plane. Trits are then ranked in descending order of $-\Delta D/\Delta R$, so that those expected to yield larger distortion reduction per bit are transmitted earlier.

\subsection{Trit Partitioning Within a Trit-Plane}
\label{ssec:trit_partition}

Using the initial scale parameters $\mathbf{\Sigma}^0$ predicted from the hyper latent $\hat{\mathbf{Z}}$, we determine the number of trits required to represent each latent element. For every $\Sten_i^0$, we compute the width of the symmetric Gaussian interval that contains probability mass $1 - 10^{-9}$. This interval length is given by $2\kappa\Sten_i^0$, where $\kappa = -\Phi^{-1}(5\times10^{-10})$. The required number of trits is then obtained by taking the ceiling of the base-3 logarithm of this length:
\begin{equation}
n_i = \lceil \log_3(2\kappa\Sten_i^0) \rceil.
\end{equation}
Thus, $n_i$ represents the number of ternary digits needed to encode the $i$-th latent element. In each trit-plane, before processing the first step, we partition all elements satisfying $n_i \ge l$ in the current plane $l$ into multiple subsets, which correspond to the coding steps.

\subsection{Algorithmic Workflow}
\label{ssec:alg_workflow}

Since the encoder and decoder share the same transmission schedule described in Section~\ref{ssec:trans_schedule}, they can identify the trits processed at each step. Once the trits for a given step are decoded, the corresponding latent bounds are updated by shrinking to the selected sub-intervals. Full pseudo-code for the encoding and decoding procedures is provided in Algorithms~\ref{alg:encoding} and~\ref{alg:decoding}. During training, the forward pass follows the same procedure as the decoding process in Algorithm~\ref{alg:decoding}, except that no entropy decoding is performed. Instead, the lower and upper bounds at each step are directly obtained from the ground-truth latent $\mathbf{Y}$.

\begin{algorithm}[hbt!]
\caption{Encoding in BLI}
\label{alg:encoding}
\KwData{Initial parameters $\Mten^0, \Sten^0$, trit counts $n_i$}
\KwResult{Encoded bitstream}

\BlankLine
\noindent{\ttfamily /* Initialization */}
\BlankLine
Compute initial bounds:
\[
\mathbf{C}^0_i \gets \Mten^0_i, \qquad
\mathbf{D}^0_i \gets 3^{n_i}.
\]

Set step index $k \gets 1$\;
\BlankLine
\While{not all trits encoded}{
  \BlankLine
  Estimate $(\Mten^k,\Sten^k, \hat{\Yten}^{k-1})$ using 
  $(\Cten^{k-1},\Dten^{k-1},\Mten^0,\Sten^0)$ 
  \tcp*{BLI}

  \BlankLine
  Identify trits to be encoded for step $k$\;
  
  \BlankLine
  Encode the trits for step $k$ using $(\Mten^k,\Sten^k)$ 
  \tcp*{Entropy encoding}
  
  \BlankLine
  Update bounds $(\Cten^k,\Dten^k)$ 
  \tcp*{Bound update}
  
  \BlankLine
  $k \gets k+1$\;
}
\end{algorithm}

\begin{algorithm}[hbt!]
\caption{Decoding in BLI}
\label{alg:decoding}
\KwData{Initial parameters $\Mten^0, \Sten^0$, trit counts $n_i$, encoded bitstream}
\KwResult{Reconstructed latent}

\BlankLine
\noindent{\ttfamily /* Initialization */}
\BlankLine
Compute initial bounds:
\[
\mathbf{C}^0_i \gets \Mten^0_i, \qquad
\mathbf{D}^0_i \gets 3^{n_i}.
\]

Set step index $k \gets 1$\;
\BlankLine
\While{target quality or bitrate not reached}{
  \BlankLine
  Estimate $(\Mten^k,\Sten^k, \hat{\Yten}^{k-1})$ using 
  $(\Cten^{k-1},\Dten^{k-1},\Mten^0,\Sten^0)$ 
  \tcp*{BLI}
  
  \BlankLine
  Identify trits to be decoded for step $k$\;

  \BlankLine
  Decode the trits for step $k$ from the bitstream using $(\Mten^k,\Sten^k)$ 
  \tcp*{Entropy decoding}

  \BlankLine
  Update bounds $(\Cten^k,\Dten^k)$ using decoded trits
  \tcp*{Bound update}

  \BlankLine
  $k \gets k + 1$\;
}
\end{algorithm}

\subsection{Baseline Implementations}
\label{ssec:baseline_impl}

For the FTIC \citep{y2024_ICLR_li} backbone, which underlies FTIC-DPICT \citep{y2022_CVPR_lee}, FTIC-CTC \citep{y2023_CVPR_jeon}, and BLI, we replaced the original entropy model with a simple structure described in Section~\ref{ssec:arch_configs} and initialized the network using the pretrained FTIC weights released in its official GitHub repository for training.

We implement the traditional codecs JPEG2000 \citep{y2001_SPM_JPEG2000}, BPG444 \citep{bpg}, and VTM 12.0 \citep{y2021_TCSVT_Bross} as follows.

\medskip
\noindent
\textbf{JPEG2000:} We employ the open-source implementation provided in \citet{y2001_SPM_JPEG2000}. The commands used for encoding and decoding are provided below. We convert the RGB images into raw format before processing.

\begin{lstlisting}
{buildpath}/opj_compress -i {inputfile}
-o {bin} -r {15:150}
-F {width},{height},3,8,u@1x1:1x1:1x1

{buildpath}/opj_decompress -i {bin} -o
{outputfile}
\end{lstlisting}

\medskip
\noindent
\textbf{BPG444:} We use the software provided in \citet{bpg} and run the commands listed below.

\begin{lstlisting}
{buildpath}/bpgenc {inputfile} -o {bin}
-q {26:52}
-f 444 -e x265

{buildpath}/bpgdec -o {outputfile} {bin}
\end{lstlisting}

\medskip
\noindent
\textbf{VTM 12.0:} We run the VTM 12.0 reference software package available at \url{https://vcgit.hhi.fraunhofer.de/jvet/VVCSoftware_VTM/-/tree/VTM-12.0} using the following commands.

\begin{lstlisting}
{buildpath}/EncoderApp -i {inputfile} -c
{cfgpath}/encoder_intra_vtm.cfg
-o /dev/null -b {bin} -wdt {width} -hgt
{height} -fr 1 -f 1
-q 24, 26, 30, 31:43
--InputChromaFormat=444
--InputBitDepth=8
--ConformanceWindowMode=1
--InputColourSpaceConvert=RGBtoGBR
--SNRInternalColourSpace=1
--OutputInternalColourSpace=0

{buildpath}/DecoderApp -b {bin} -o
{outputfile} -d 8
--OutputColourSpaceConvert=GBRtoRGB
\end{lstlisting}

For the other methods in comparison, we use their official implementations or the reported results provided in their papers.

\section{Extended Analyses and Results}
\label{sec:ext_results}

\subsection{Computational Complexity}
\label{ssec:comp_complexity}

To further analyze the computational cost of the progressive codecs, we report the floating-point operations and decompose the end-to-end latency into neural-network inference and non-neural processing. The FLOP counts are measured for a $768\times512$ input, corresponding to the Kodak resolution. Latency is measured on the Kodak dataset using an AMD Ryzen 9 3900X CPU and an NVIDIA RTX 4090 GPU. The non-neural processing time includes probability processing and entropy coding outside the neural-network forward passes.

\begin{table}[t]
\centering
\caption{Comparison of computational complexity on the Kodak dataset.}
\label{table:comp_complexity}
\begin{tabular}{lcccc}
\toprule
Method & FLOPs (T) & Neural inference (s) & Non-neural processing (s) & Total latency (s) \\
\midrule
FTIC-DPICT & 41.42 & 0.08 & 5.77 & 5.85 \\
FTIC-CTC & 66.69 & 0.18 & 6.44 & 6.62 \\
BLI & 65.96 & 3.72 & 1.05 & \textbf{4.77} \\
\bottomrule
\end{tabular}
\end{table}

BLI requires more neural computation than FTIC-DPICT because it repeatedly refines the entropy distribution and latent estimate through in-plane autoregressive inference. Its FLOP count is comparable to that of FTIC-CTC, while its neural inference time is substantially higher than those of both FTIC-DPICT and FTIC-CTC. Nevertheless, BLI achieves the lowest end-to-end latency because its non-neural processing overhead is significantly smaller.

This difference mainly arises from the computation of trit probabilities. FTIC-DPICT and FTIC-CTC construct progressive discrete probabilities using loop-based probability composition, whereas BLI directly obtains the ternary probabilities by integrating the predicted Gaussian distribution over the corresponding sub-intervals. Consequently, the additional neural computation introduced by the BLI refinement steps is offset by substantially lower non-neural processing overhead, resulting in the lowest overall latency among the three progressive codecs.

\subsection{Sensitivity to Transmission Ordering}
\label{ssec:transmission_order}

We further analyze the effect of the within-plane transmission ordering on the progressive R-D performance of BLI. Using the same trained BLI model, we compare the R-D-prioritized ordering used in our main experiments with three alternative strategies: random ordering, raster scan, and ascending order of the predicted scale parameter. Table~\ref{table:transmission_order} reports two types of BD-rate. The full BD-rate is computed using all progressive reconstruction points, whereas the endpoint BD-rate is computed using only the fully decoded trit-plane endpoints.

\begin{table}[t]
\centering
\caption{Effect of the within-plane transmission ordering on the Kodak dataset. BD-rates (\%) are measured relative to FTIC-DPICT.}
\label{table:transmission_order}
\begin{tabular}{lcc}
\toprule
Trans. ordering & Full BD-rate (\%) & Endpoint BD-rate (\%) \\
\midrule
Random & $-10.01$ & $-19.63$ \\
Raster scan & $-7.62$  & $-18.52$ \\
$\sigma$ ascending & $-2.47$  & $-18.02$ \\
R-D prioritized & $\mathbf{-17.78}$ & $-19.59$ \\
\bottomrule
\end{tabular}
\end{table}

The transmission ordering has a substantial effect on the full progressive R-D performance, while its effect on the trit-plane endpoints is relatively small. The results indicate that transmission ordering mainly affects how efficiently information is delivered at intermediate prefixes within each trit-plane, while the reconstruction quality at fully decoded trit-plane endpoints is less sensitive to the ordering. The R-D-prioritized strategy therefore improves progressive efficiency primarily by transmitting high-impact trits earlier within each plane.

\subsection{Entropy Modeling vs. Latent Reconstruction}
\label{ssec:entropy_recon_ablation}

To assess the relative contributions of entropy modeling and latent reconstruction, we perform test-time ablations using the same trained BLI model. We selectively disable each refinement pathway while retaining the other. Table~\ref{table:entropy_recon_ablation} summarizes the results on the Kodak dataset.

\begin{table}[t]
\centering
\caption{Test-time ablation of entropy modeling and latent reconstruction on the Kodak dataset. BD-rates (\%) are measured relative to FTIC-DPICT.}
\label{table:entropy_recon_ablation}
\begin{tabular}{lccc}
\toprule
Test-time variant & Entropy params. for coding & Latent recon. & BD-rate (\%) \\
\midrule
Recon. with Gaussian mean & $(\Mten^{k}, \Sten^{k})$ & Gaussian mean & $-16.38$ \\
No entropy refinement & $(\Mten^{0}, \Sten^{0})$ & $\hat{\Yten}^{k-1}$ & $-10.24$ \\
Full BLI & $(\Mten^{k}, \Sten^{k})$ & $\hat{\Yten}^{k-1}$ & $\mathbf{-17.78}$ \\
\bottomrule
\end{tabular}
\end{table}

When the step-wise refined entropy parameters $(\mathbf{M}^{k}, \mathbf{\Sigma}^{k})$ are retained but the predicted latent estimate $\hat{\mathbf{Y}}^{k-1}$ is replaced by the Gaussian expectation, the BD-rate changes from $-17.78\%$ to $-16.38\%$. In contrast, disabling entropy refinement and using the initial hyperprior parameters $(\mathbf{M}^{0}, \mathbf{\Sigma}^{0})$ for entropy coding, while retaining the predicted latent reconstruction, reduces the BD-rate gain to $-10.24\%$. These results indicate that progressively refining the entropy parameters provides a larger contribution to the overall R-D improvement. The explicit latent-reconstruction prediction provides a smaller but consistent additional gain, showing that the two pathways play complementary roles in BLI.

\subsection{Perceptual Evaluation}

We further evaluate perceptual quality at low bitrates on the Kodak dataset using LPIPS and DISTS. For each method, the metric values at common target bitrates of 0.10, 0.15, and 0.20~bpp are obtained by linear interpolation between adjacent measured operating points. Table~\ref{table:perceptual_quality} reports the dataset-level averages over all Kodak images, where lower values indicate better perceptual quality.

\begin{table}[t]
\centering
\caption{Comparison of perceptual quality on the Kodak dataset at low bitrates. Lower LPIPS and DISTS values indicate better perceptual quality.}
\label{table:perceptual_quality}
\begin{tabular}{llrrrrr}
\toprule
bpp & Metric & BPG444 & VTM 12.0 & FTIC-DPICT & FTIC-CTC & BLI \\
\midrule
0.10 & LPIPS & 0.3848 & 0.3494 & 0.4008 & 0.3465 & \textbf{0.3203} \\
     & DISTS & 0.2038 & 0.1994 & 0.2308 & 0.2089 & \textbf{0.1910} \\
\midrule
0.15 & LPIPS & 0.3207 & 0.3064 & 0.3569 & 0.3123 & \textbf{0.2868} \\
     & DISTS & 0.1771 & 0.1800 & 0.2049 & 0.1884 & \textbf{0.1736} \\
\midrule
0.20 & LPIPS & 0.2760 & 0.2663 & 0.3045 & 0.2661 & \textbf{0.2366} \\
     & DISTS & 0.1586 & 0.1606 & 0.1780 & 0.1630 & \textbf{0.1510} \\
\bottomrule
\end{tabular}
\end{table}

BLI achieves the lowest LPIPS and DISTS values at all three target bitrates, consistently outperforming both conventional codecs and the progressive learned baselines. These results indicate that the R-D improvements of BLI are also accompanied by improved perceptual quality in the low-bitrate regime, rather than being limited to distortion metrics such as PSNR and MS-SSIM.

\subsection{Reconstructed Images}
\label{ssec:rec_images}

\begin{figure*}[t]
\centering
\includegraphics[width=\linewidth]{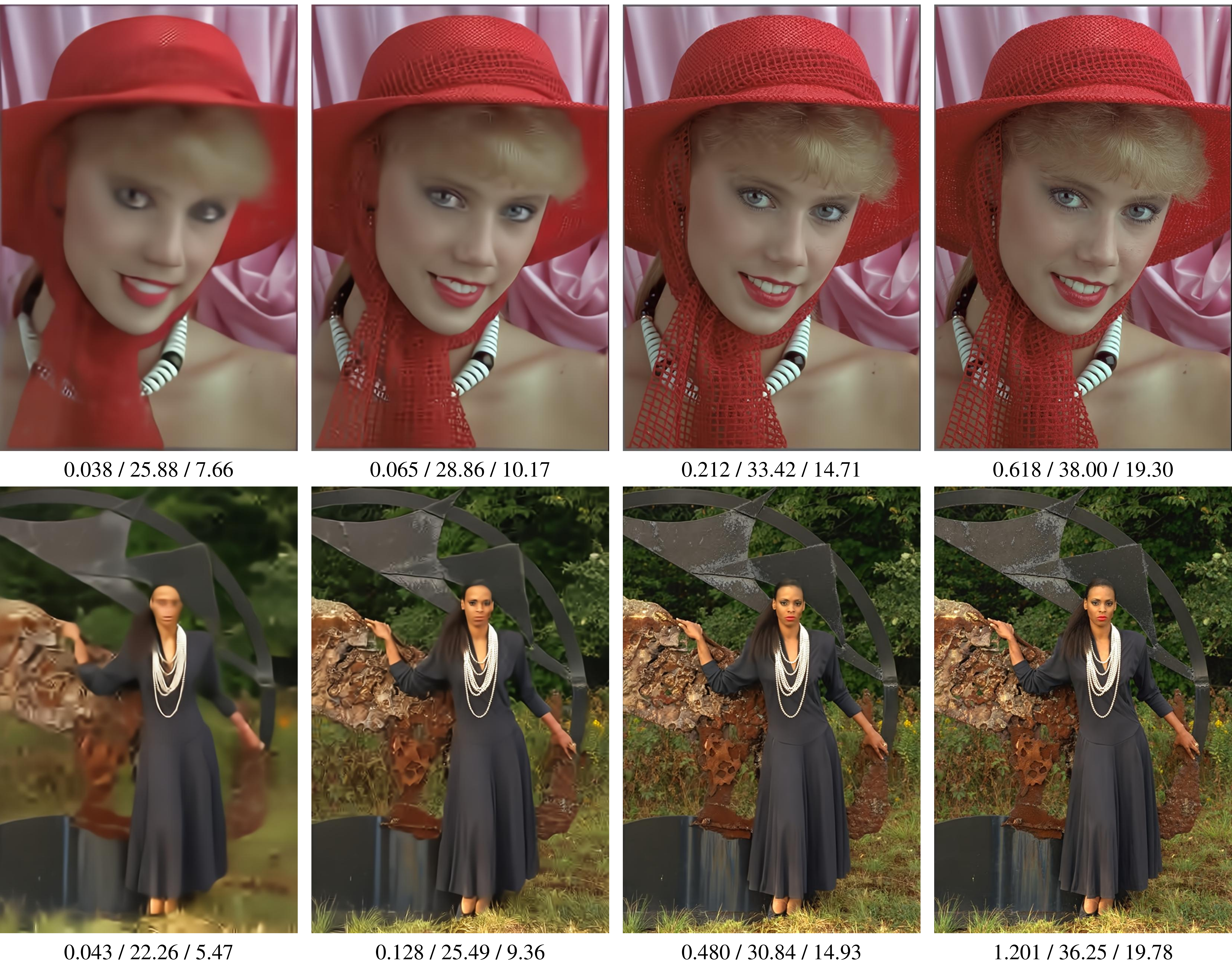}
\setlength{\abovecaptionskip}{-0.2cm}
\caption{Progressively reconstructed images of `kodim04.png' and `kodim18.png' obtained by the PSNR-trained model. The bitrate (bpp), PSNR (dB), and MS-SSIM (dB) are reported below each image.}
\label{fig:appendix_kodim_1}
\vspace*{-0.2cm}
\end{figure*}

\begin{figure*}[t]
\centering
\includegraphics[width=\linewidth]{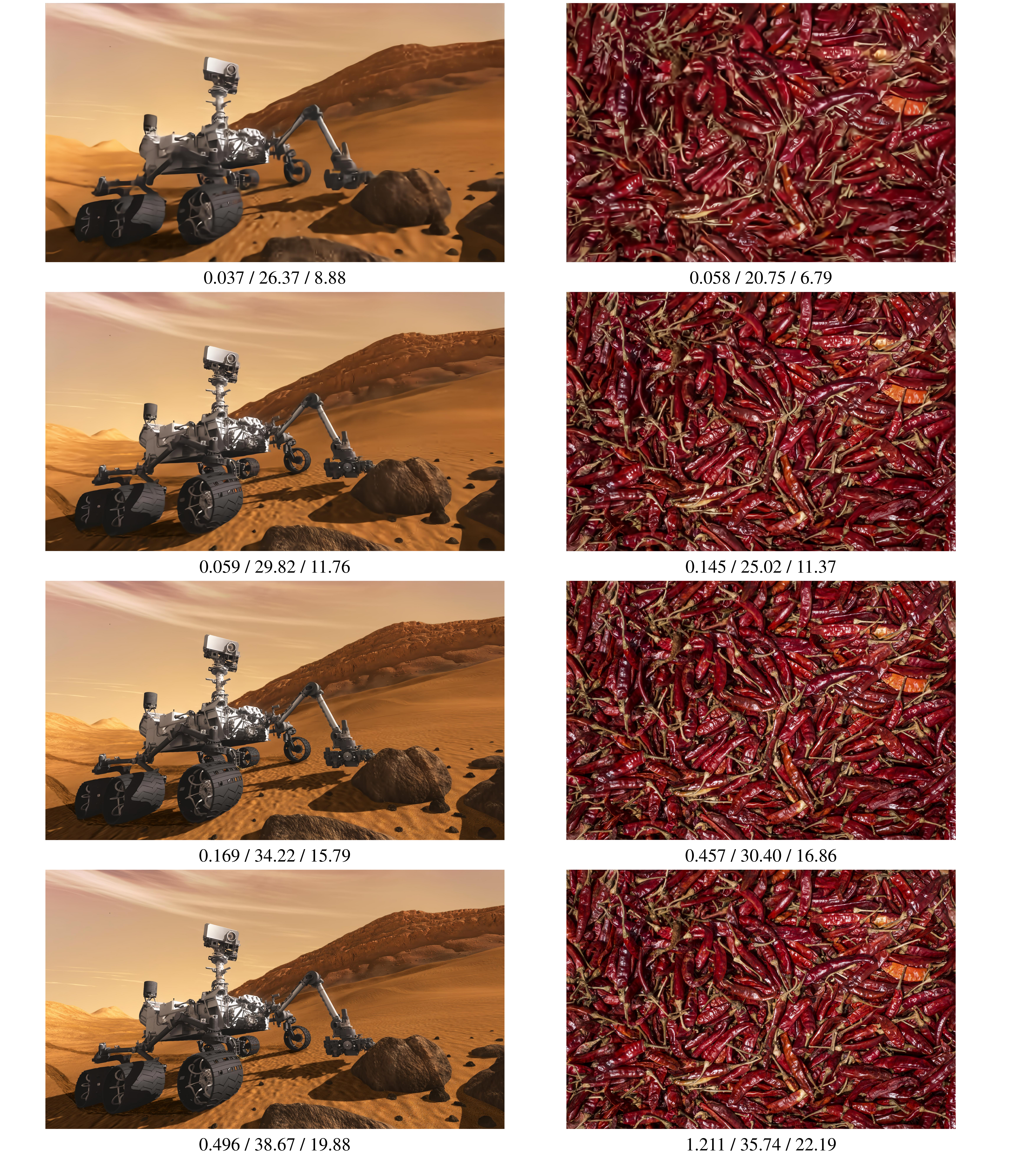}
\setlength{\abovecaptionskip}{-0.2cm}
\caption{Progressively reconstructed images of `00010\_TE\_2000x1128.png' and `00011\_TE\_1744x1160.png' obtained by the PSNR-trained model. The bitrate (bpp), PSNR (dB), and MS-SSIM (dB) are reported below each image.}
\label{fig:appendix_jpegai_3}
\vspace*{-0.2cm}
\end{figure*}

Figures~\ref{fig:appendix_kodim_1}$\sim$\ref{fig:appendix_jpegai_3} present examples of images progressively reconstructed by the PSNR-trained BLI model on the Kodak \citep{kodak}, CLIC \citep{clic2021}, and JPEG-AI \citep{jpegai} datasets.

\subsection{Tests on Challenging Color Patterns}
\label{ssec:test_highfreq}

\begin{figure*}[t]
\centering
\includegraphics[width=\linewidth]{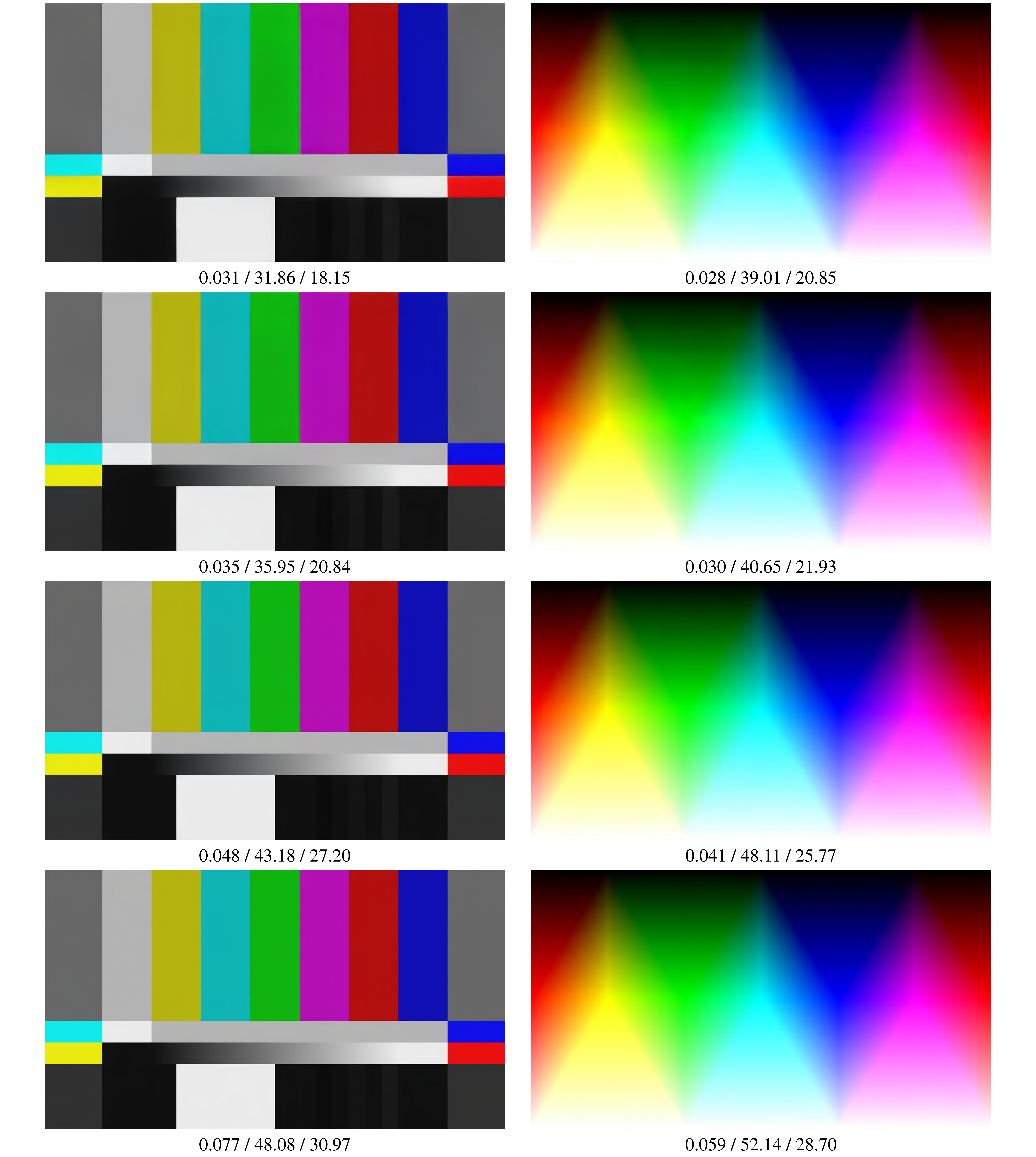}
\setlength{\abovecaptionskip}{-0.2cm}
\caption{Progressively reconstructed images of `img\_1280x720\_3x8bit\_RGB\_color\_SMPTE\_RP\_219\_2002.png' and `img\_1280x720\_3x8bit\_RGB\_color\_rainbow\_gradient.png' obtained by the PSNR-trained model. The bitrate (bpp), PSNR (dB), and MS-SSIM (dB) are reported below each image.}
\label{fig:appendix_testimages}
\vspace*{-0.2cm}
\end{figure*}

To evaluate the robustness of BLI on challenging color patterns, we apply our framework to two images from the Tecnick dataset \citep{y2014_STAG_asuni} that includes smooth low-frequency gradients and sharp chromatic edges. The resulting reconstructions are shown in Figure~\ref{fig:appendix_testimages}.

\clearpage

\begin{figure*}[t]
\centering
\includegraphics[width=\linewidth]{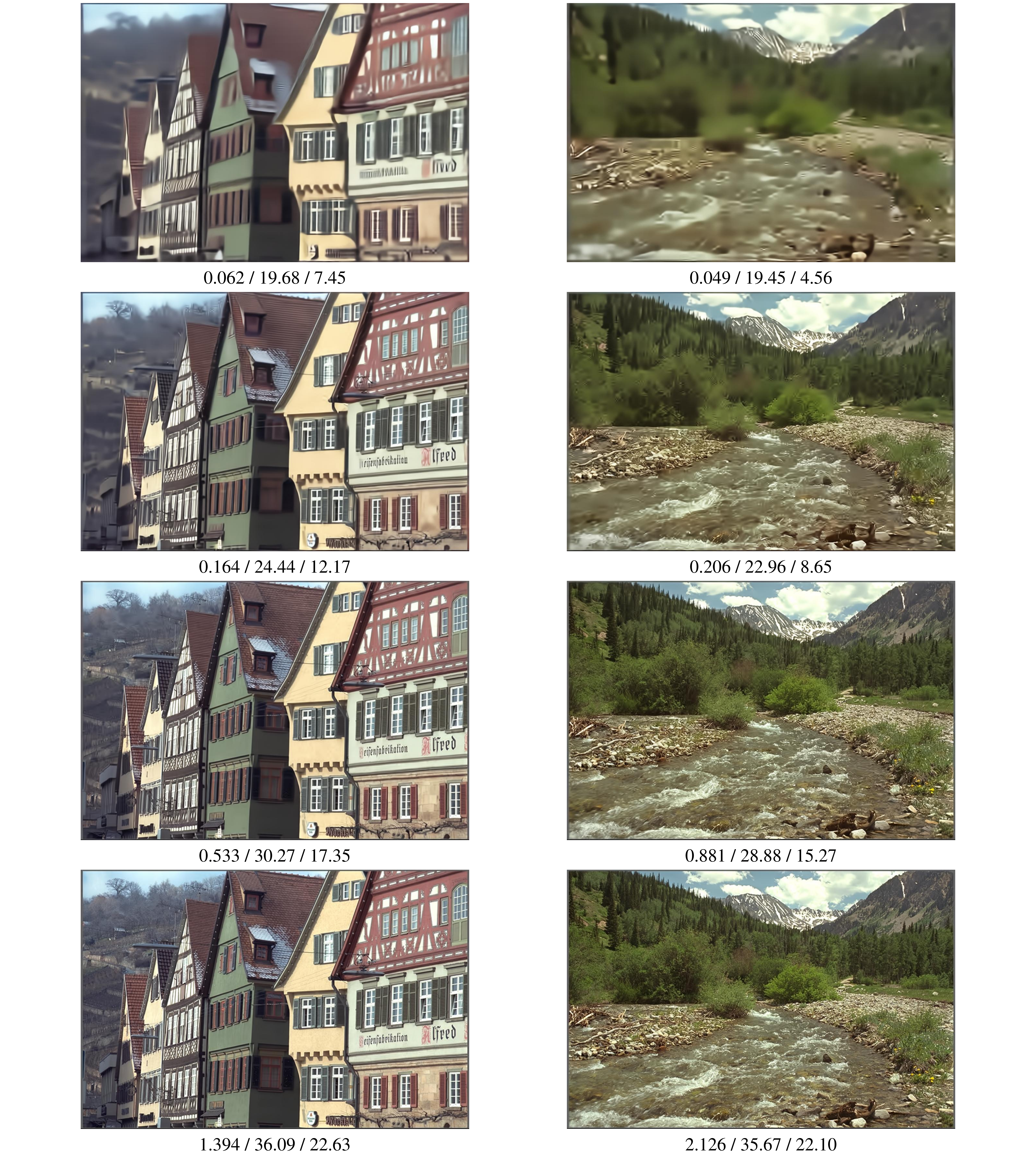}
\setlength{\abovecaptionskip}{-0.2cm}
\caption{Progressively reconstructed images of `kodim08.png' and `kodim13.png' obtained by the PSNR-trained model. The bitrate (bpp), PSNR (dB), and MS-SSIM (dB) are reported below each image.}
\label{fig:appendix_kodim_2}
\vspace*{-0.2cm}
\end{figure*}

\begin{figure*}[t]
\centering
\includegraphics[width=\linewidth]{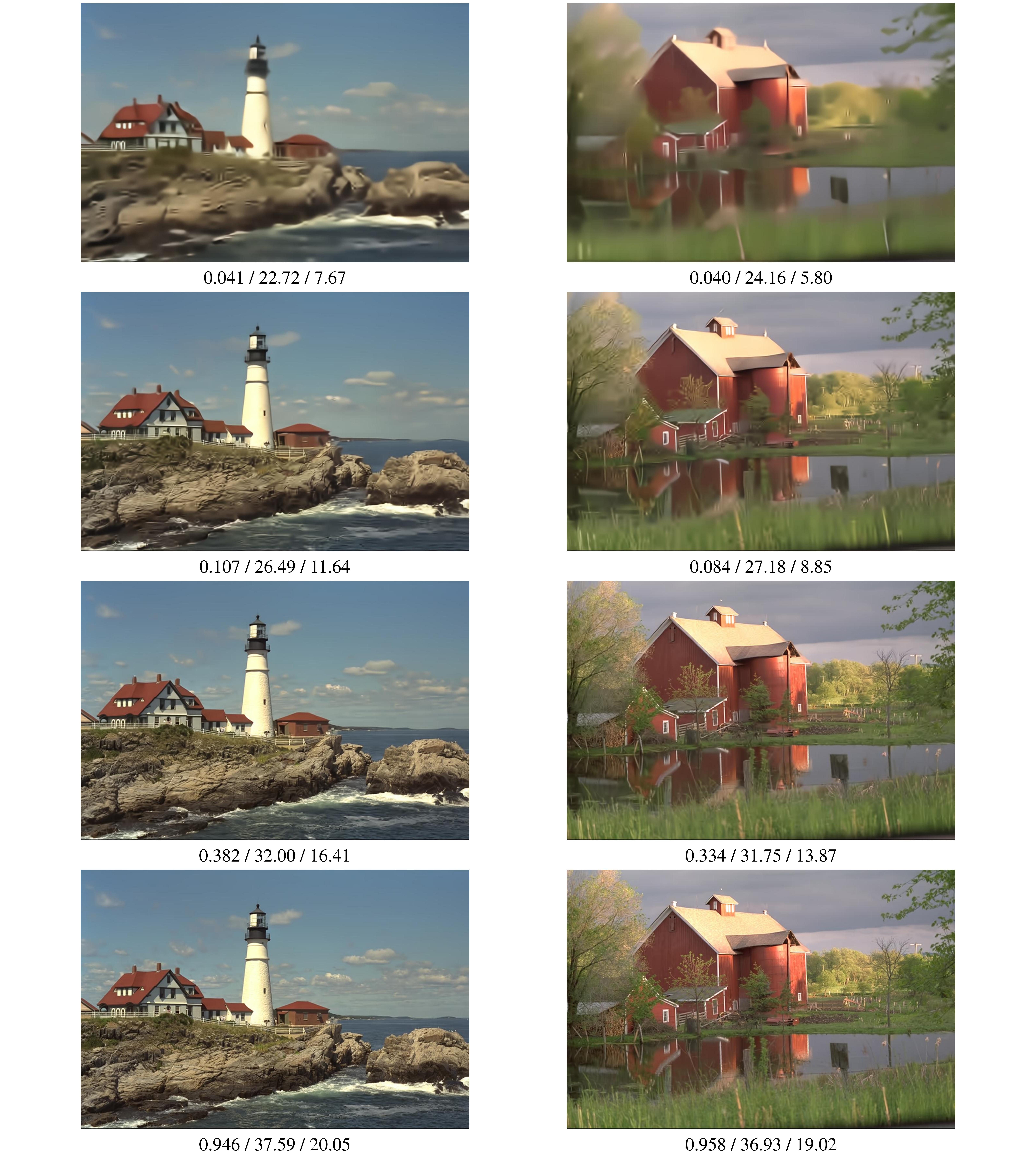}
\setlength{\abovecaptionskip}{-0.2cm}
\caption{Progressively reconstructed images of `kodim21.png' and `kodim22.png' obtained by the PSNR-trained model. The bitrate (bpp), PSNR (dB), and MS-SSIM (dB) are reported below each image.}
\label{fig:appendix_kodim_3}
\vspace*{-0.2cm}
\end{figure*}

\begin{figure*}[t]
\centering
\includegraphics[width=\linewidth]{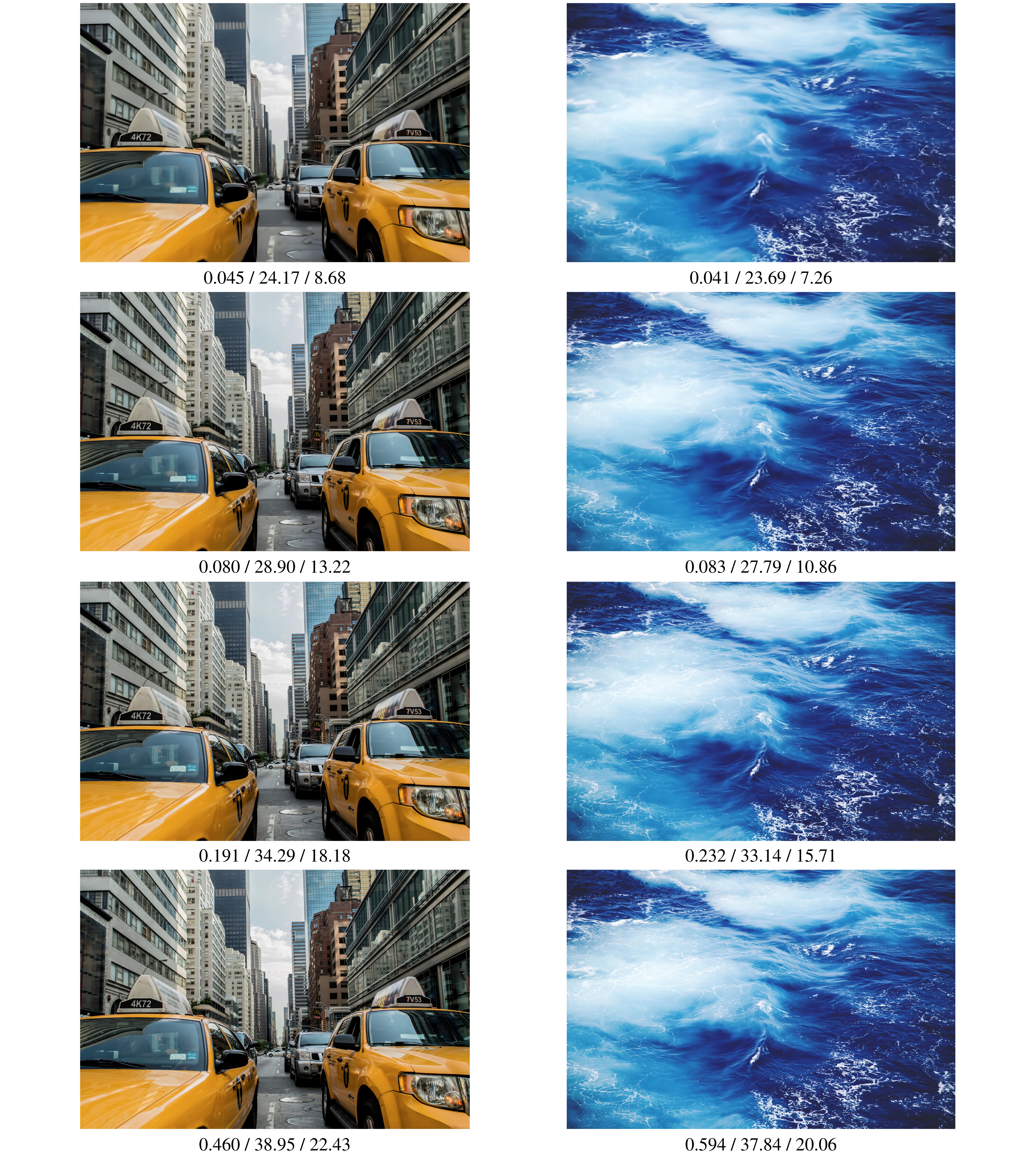}
\setlength{\abovecaptionskip}{-0.2cm}
\caption{Progressively reconstructed images of `andrew-ruiz-376.png' and `clem-onojeghuo-33741.png' obtained by the PSNR-trained model. The bitrate (bpp), PSNR (dB), and MS-SSIM (dB) are reported below each image.}
\label{fig:appendix_clic_1}
\vspace*{-0.2cm}
\end{figure*}

\begin{figure*}[t]
\centering
\includegraphics[width=\linewidth]{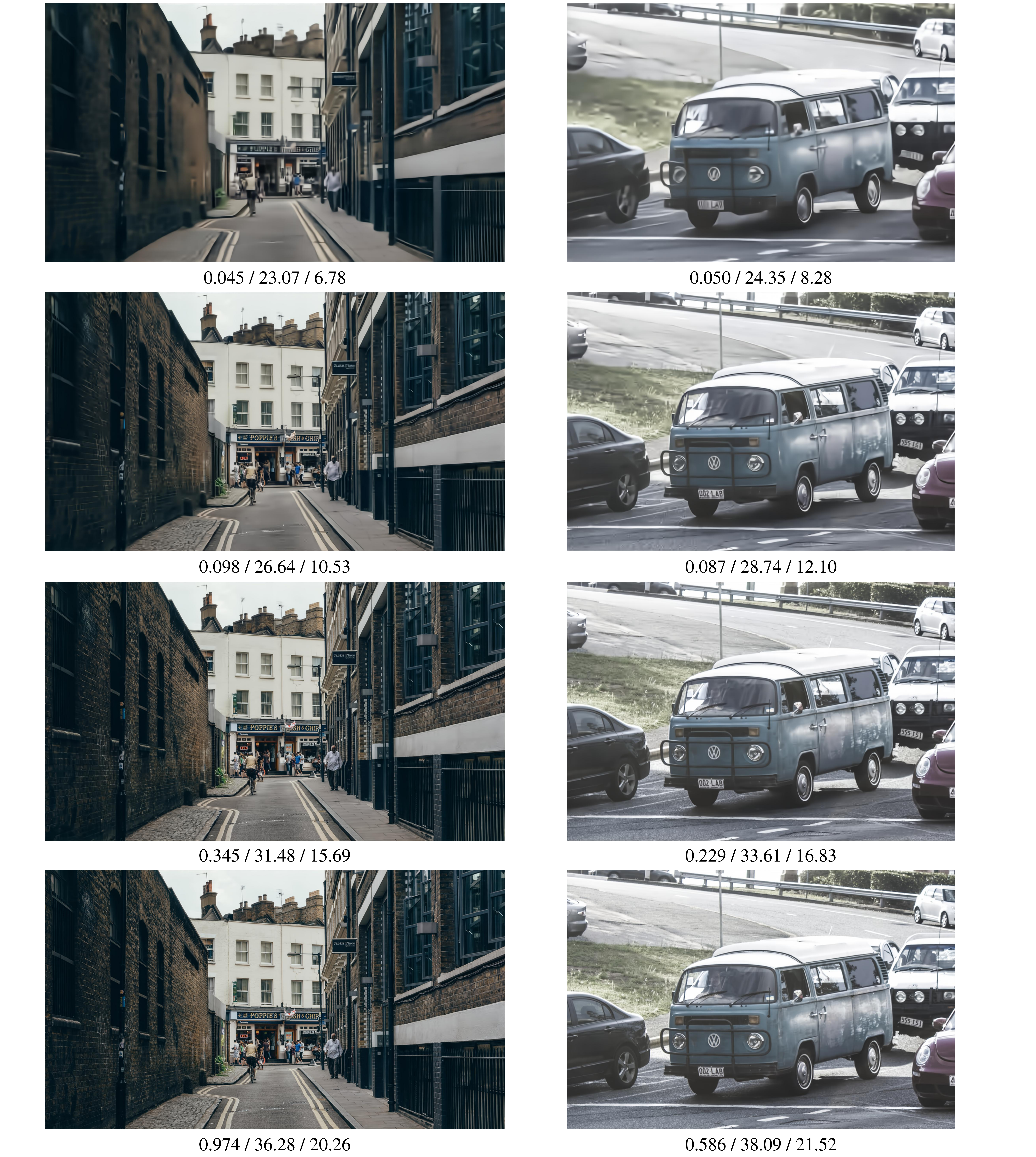}
\setlength{\abovecaptionskip}{-0.2cm}
\caption{Progressively reconstructed images of `daniel-robert-405.png' and `mullermarc-184.png' obtained by the PSNR-trained model. The bitrate (bpp), PSNR (dB), and MS-SSIM (dB) are reported below each image.}
\label{fig:appendix_clic_2}
\vspace*{-0.2cm}
\end{figure*}

\begin{figure*}[t]
\centering
\includegraphics[width=\linewidth]{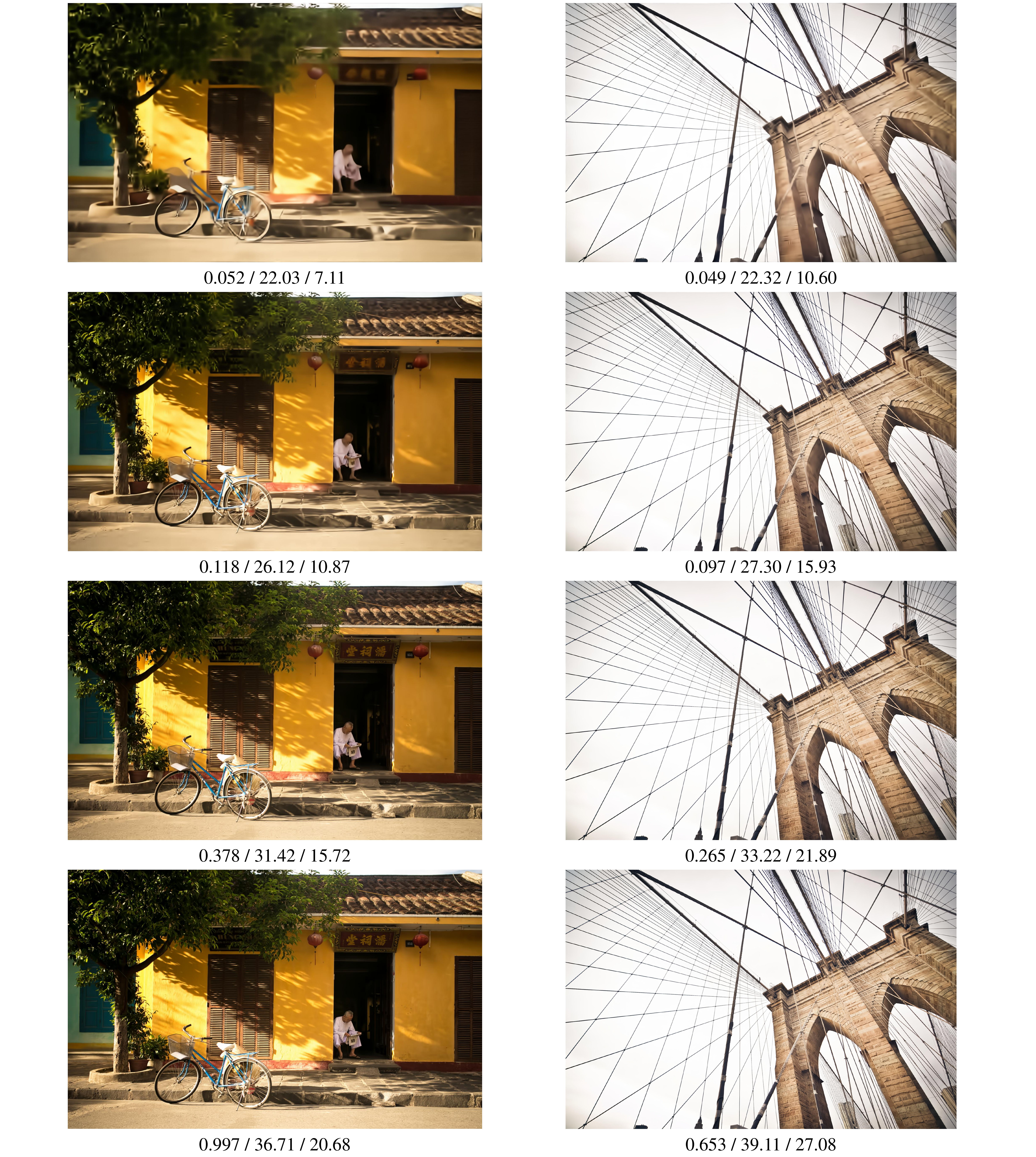}
\setlength{\abovecaptionskip}{-0.2cm}
\caption{Progressively reconstructed images of `thong-vo-428.png' and `vita-vilcina-3055.png' obtained by the PSNR-trained model. The bitrate (bpp), PSNR (dB), and MS-SSIM (dB) are reported below each image.}
\label{fig:appendix_clic_3}
\vspace*{-0.2cm}
\end{figure*}

\begin{figure*}[t]
\centering
\includegraphics[width=\linewidth]{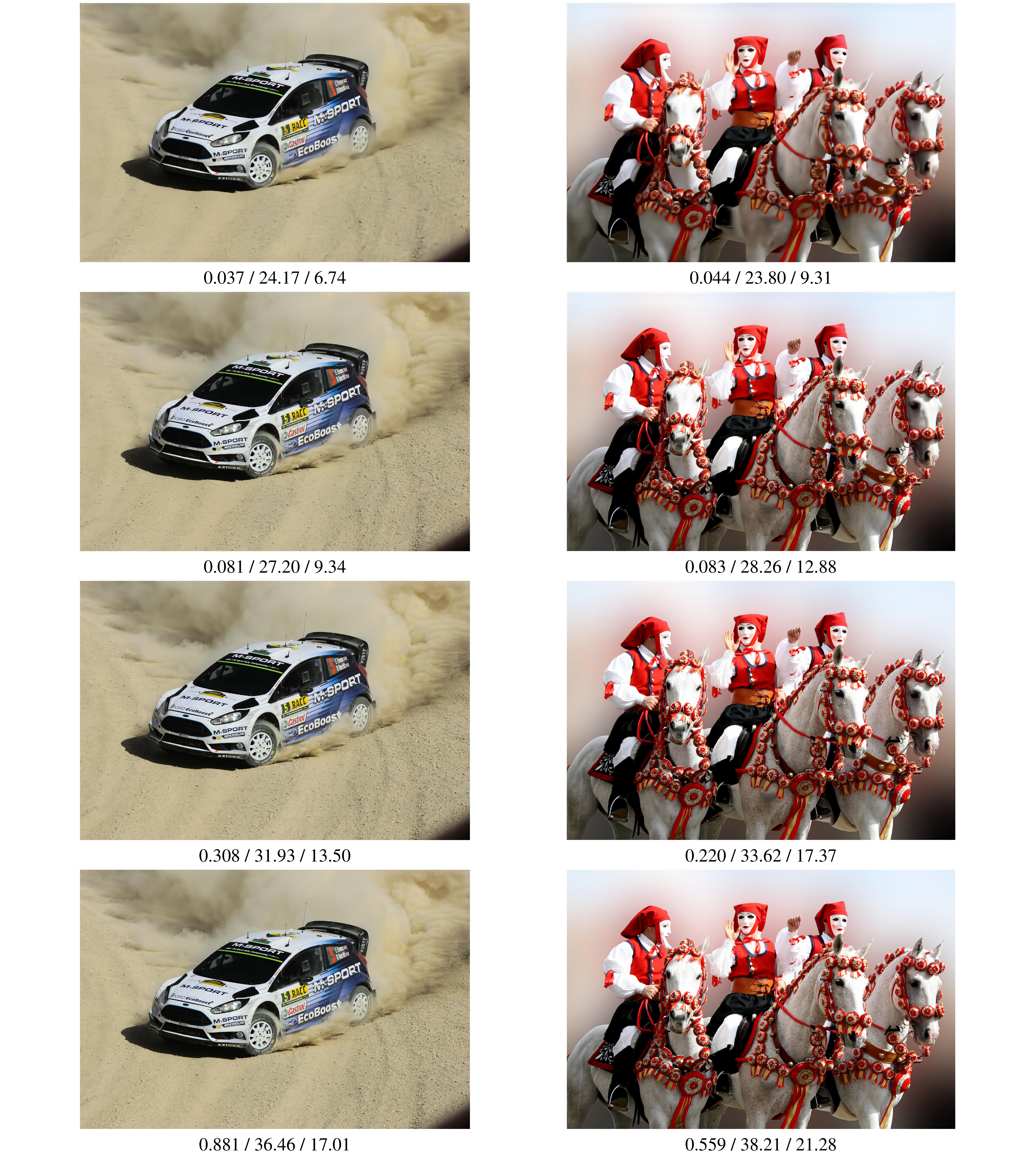}
\setlength{\abovecaptionskip}{-0.2cm}
\caption{Progressively reconstructed images of `00002\_TE\_2144x1424.png' and `00003\_TE\_1944x1296.png' obtained by the PSNR-trained model. The bitrate (bpp), PSNR (dB), and MS-SSIM (dB) are reported below each image.}
\label{fig:appendix_jpegai_1}
\vspace*{-0.2cm}
\end{figure*}

\begin{figure*}[t]
\centering
\includegraphics[width=\linewidth]{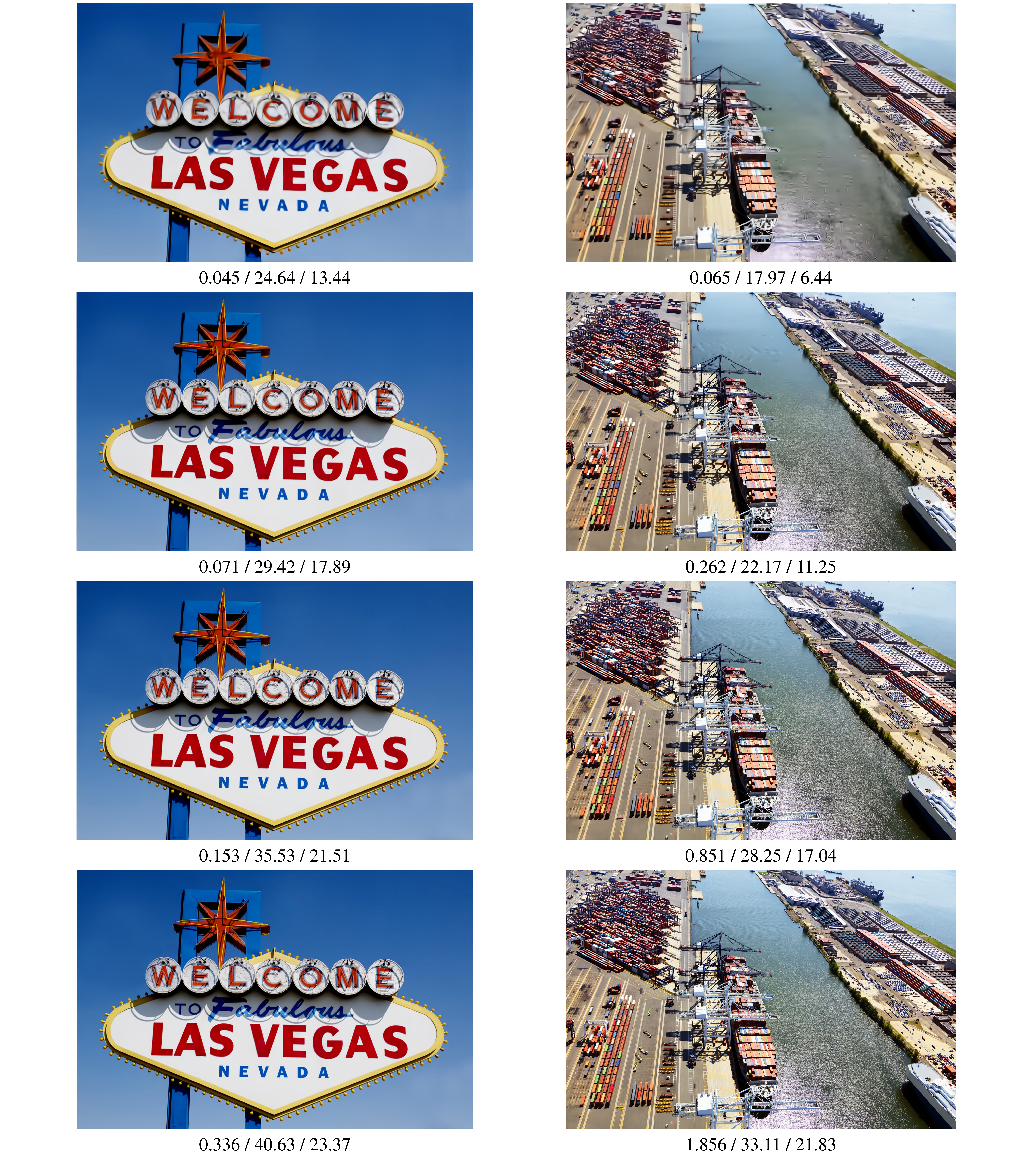}
\setlength{\abovecaptionskip}{-0.2cm}
\caption{Progressively reconstructed images of `00005\_TE\_1336x872.png' and `00009\_TE\_1976x1312.png' obtained by the PSNR-trained model. The bitrate (bpp), PSNR (dB), and MS-SSIM (dB) are reported below each image.}
\label{fig:appendix_jpegai_2}
\vspace*{-0.2cm}
\end{figure*}

\clearpage

\end{document}